\documentclass[twocolumn,tighten, times ]{aastex701}
\submitjournal{ApJS}

\newcommand{\teff}{T$_{\rm eff}$ }
\newcommand{\logg}{$\rm log\,g$ }
\newcommand{\feh}{[Fe/H] }
\newcommand{\vmic}{$v_{\rm mic}$ }
\newcommand{\vmac}{$v_{\rm mac}$ }
\newcommand{\vsini}{$v\sin i$}
\usepackage{sidecap}
\usepackage{multirow}
\usepackage{rotating}
\usepackage{subcaption}
\usepackage{natbib}
\begin{document}

\title{Stellar parameters and abundances for 3.2 million DESI DR1 spectra using machine learning}

\author[orcid=0000-0003-2947-5889]{Vernekar Nagaraj}
\affiliation{Homer L. Dodge Department of Physics and Astronomy, University of Oklahoma, 440 W. Brooks St., Norman, OK, 73019 USA}
\email[show]{nv101@ou.edu}  

\author[orcid=0000-0001-7294-9766]{Michael Hayden} 
\affiliation{Homer L. Dodge Department of Physics and Astronomy, University of Oklahoma, 440 W. Brooks St., Norman, OK, 73019 USA}
\email{mrhayden@ou.edu}

\author[orcid=0000-0001-8808-0073]{Sara Lucatello}
\affiliation{INAF - Ossevatorio Astronomico di Padova, vicolo dell'Osservatorio 5, 35122 Padova, Italy}
\email{sara.lucatello@inaf.it}

\begin{abstract}
We present a machine-learning pipeline for determining stellar parameters and chemical abundances from spectra observed by the Dark Energy Spectroscopic Instrument (DESI). Building on the model-driven Payne framework, we train an artificial neural network on $\sim$370,000 synthetic spectra spanning FGK stars, and fit the network directly to the flux-calibrated, un-normalised spectral energy distribution (SED) of each star, rather than to a continuum-normalised spectrum as done by DESI's official Stellar Parameter (SP) pipeline. Internal accuracy tests show a median interpolation error of less than 0.3\% for 90\% of a synthetic verification sample, and the full fitting workflow recovers input labels to high accuracy even at low signal-to-noise ratio (S/N). Validating our method on cross-matched samples of 6719 APOGEE and 3455 GALAH stars observed by DESI with $\mathrm{S/N}\geq20$, we recover $T_{\rm eff}$ and $\log g$ with smaller systematic offsets than the DESI SP pipeline, an improvement we attribute to fitting the full SED shape rather than a normalised spectrum. We further recover 12 elemental abundances (Na, Mg, Al, Si, Ca, Ti, V, Cr, Mn, Ni, Ba, and Y) to an accuracy that matches or exceeds SP where a comparison is possible, and provide four abundances, Mn, V, Ba, and Y, that SP does not report at all. Applying this pipeline to the DESI Data Release 1 stellar sample, we deliver a catalogue of stellar parameters and abundances for 3.2 million stars.
\end{abstract}



\section{Introduction} 

The chemical and dynamical fingerprints imprinted on stellar atmospheres are among the few direct records of how the Milky Way how the Milky Way assembled and evolved over cosmic time \citep{2002ARA&A..40..487F}. Recovering these fingerprints for enough stars to trace the Galaxy's
formation history, rather than a handful of well-studied benchmarks, has been the driving motivation behind the last two decades of spectroscopic survey design. At the high-resolution end, surveys such as APOGEE\footnote{APO Galactic Evolution Experiment} ($R \sim 22{,}500$; \citealt{apogee2017}), GALAH\footnote{Galactic Archaeology with HERMES} ($R \sim 28{,}000$; \citealt{galah2015}), and Gaia-ESO\footnote{Gaia-European Southern Observatory} Survey ($R$ up to $\sim 47{,}000$; \citealt{gaiaeso2012,gaiaeso2013}), deliver detailed multi-element abundances for relatively bright samples. Complementing these, medium- and low-resolution surveys trade spectral resolution for depth and sky coverage: RAVE\footnote{The Radial Velocity Experiment} ($R \sim 7500$; \citealt{rave2006}), LAMOST\footnote{Large Sky Area Multi-Object Fiber Spectroscopic Telescope} ($R \sim 1800$; \citealt{2012RAA....12.1197C}), SEGUE-2\footnote{Sloan Extension for Galactic Understanding and Exploration} ($R \sim 1800$; \citealt{2022ApJS..259...60R}) and now the Dark Energy Spectroscopic Instrument (DESI; $R \sim 2000$--$5500$; \citealt{cooper2023}) reach fainter, and more distant stars at the cost of blended spectral features. At the extreme low-resolution end, the Gaia mission's BP/RP slitless spectrograph \citep{GaiaCollaboration2016} has already delivered spectra for 220 million stars at $R \sim 50$ in its DR3 \citep{GaiaCollaboration2021}. 
Further facilities are on the horizon: 4MOST\footnote{4-metre Multi-Object Spectroscopic Telescope} \citep{deJong2019}, WEAVE\footnote{William Hershel Telescope Enhanced Area Velocity Explorer} \citep{2024MNRAS.530.2688J} and MOONS\footnote{ Multi Object Optical and Near-infrared Spectrograph} \citep{Cirasuolo2020}. Each is expected to add millions more stellar spectra over the coming years. All three surveys offer high- and low-resolution modes, but most of the data will be low-resolution. This makes the low-resolution regime increasingly important, especially the challenge of extracting reliable parameters and abundances from these spectra. The DESI Milky Way Survey \citep{cooper2023} alone is expected to observe several million stars over the lifetime of the survey.

Traditional spectroscopic methods measure individual absorption lines or fit synthetic spectra one parameter at a time. These methods do not scale well to millions of spectra due to the massive computational power needed. Because of this, machine-learning methods have become standard for large surveys. The Cannon \citep{melissa2015} was an early data-driven approach: it trains a model directly on observed stars with known labels. StarNet \citep{starnet} adopts a similar data-driven approach using convolutional neural networks. Payne \citep{Payne} took a different, model-driven approach of deriving stellar parameters and abundances. It trains an artificial neural network (ANN) on synthetic spectra, so the network acts as a fast spectral interpolator. Observed spectra are then fit to this interpolator using standard $\chi^{2}$ minimisation. The main advantage of these methods over the traditional ones is speed. A single-star analysis that used to take about an hour in the traditional method takes less than 5 sec with the machine learning algorithms. 

DESI provides a catalogue with results from two official stellar-parameter pipelines: the RVS pipeline, which uses template cross-correlation, and the Stellar Parameter (SP) pipeline, which is based on FERRE \citep{koposov2025}. Like most large-survey pipelines, both work on continuum-normalised spectra. This means the pseudo-continuum is removed first, and only the remaining line signal is fit. \citet{zhang2024} conducted an independent analysis using a data-driven approach instead. They applied a variant of the Payne (the Data-Driven Payne) directly to DESI Early Data Release spectra (roughly 500k spectra), recovering 15 labels including abundances for 12 elements, again after normalisation. In an earlier paper \citep[LRPayne;][]{vernekar2025}, we applied a similar model-driven ANN to continuum-normalised, low-resolution spectra ($R\sim5000$, 4200-6900\,\AA) representative of the WEAVE survey. We validated it using 64 benchmark and metal-poor stars.

In this paper, we take a different approach for DESI. We use the neural network to fit the un-normalised spectrum directly. Continuum normalisation is standard, but it discards meaningful data. The overall shape of the spectrum's energy distribution tells us about temperature and interstellar reddening, information that is particularly important at low/intermediate resolution, where a limited number of spectral features can be cleanly measured. Normalising removes that signal. The normalising procedure also adds its own errors, which are hard to assess, especially at DESI's moderate signal-to-noise. So instead, we fit the flux-calibrated spectrum directly. This lets us test whether we recover extra information compared to DESI's official pipelines. 

The paper is organised as follows. Section 2 describes the neural network architecture, the training data, the pixel masking, and how we fit the observed DESI spectra. Section 3 presents our results: we choose the best network among the architectures we tested, then verify our stellar parameters and abundances against APOGEE and GALAH, including a comparison with DESI's own official pipeline. Section 4 summarises our findings.

\section{Method}

\subsection{Architecture}

For the algorithm, we adopt the same model-driven approach used in LRPayne\footnote{\url{https://github.com/nagarajvernekar30/LRPayne}} \citep{vernekar2025}, itself based on The Payne\footnote{\url{https://github.com/tingyuansen/The_Payne}} \citep{Payne}, with modifications suited to fitting flux-calibrated, un-normalised DESI spectra. At its core, the method uses a fully connected artificial neural network (ANN) as a spectral interpolator, predicting the flux at each wavelength point given a set of stellar labels. This allows the network to simultaneously fit multi-dimensional labels using the whole spectrum, capturing the covariances between labels caused by line blending at low resolution.

As in LRPayne, we use a five-layer network with three hidden layers. We tested hidden-layer widths of 200, 300, 500, 1000, and 2000 neurons. If only the stellar parameters (T$_{\rm eff}$, log\,g, [Fe/H]) were required, a width of 500 neurons was sufficient to achieve good performance. However, since we aimed to also recover individual elemental abundances, a larger network was needed to accurately capture the more subtle spectral features associated with them. We found that 2000 neurons per hidden layer gave the best overall performance for both parameters and abundances. We therefore adopt this width for our network. Between the first four layers, we use a LeakyReLU activation function ($\mathrm{LeakyReLU}(x) = \max(0,x) + \mathrm{slope} \times \min(0,x)$), which helps prevent the vanishing-gradient problem that can otherwise slow or stall convergence. Unlike LRPayne, where a sigmoid activation was used between the last hidden layer and the output layer to constrain the output to the [0,1] range appropriate for a normalised spectrum, here we use LeakyReLU for the output layer as well, since the network must reproduce the flux-calibrated, un-normalised spectral energy distribution, whose values are not bounded to this range. The network is built using TensorFlow \citep{tensorflow2015-whitepaper}. For a pictorial representation of the workflow and the architecture of the ANN, refer to Figure 1 of \citep{vernekar2025}.

\subsection{Synthetic Spectra} \label{sec:synthetic}

The synthetic spectral library was generated using 1D MARCS\footnote{Model Atmospheres with a Radiative and Convective Scheme} model atmospheres \citep{marcs} and the Turbospectrum synthesis code \citep{1998A&A...330.1109A,turbospectrum} within iSpec \citep{ispec2014,ispec2019}. The model atmospheres are 1D plane-parallel models computed under the assumption of local thermodynamic equilibrium (LTE). The atomic linelist used for the synthesis was taken from VALD\footnote{Vienna Atomic Line Database}, covering the wavelength range of the DESI spectra.

Unlike LRPayne, where spectra were synthesised in normalised form, here we generate the raw, un-normalised synthetic SED for each star. Each spectrum is synthesised at high resolution ($R=300\,000$) and then convolved down to $R=10\,000$ before being resampled onto a uniform wavelength grid with a step of 0.01\,nm. This intermediate spectrum is subsequently convolved with the DESI line-spread function (LSF) to match the instrumental resolution of the observed spectra. We adopt this two-step convolution strategy for computational reasons: convolving directly from $R=300\,000$ (around 590\,000 wavelength pixels) to the DESI resolution using a variable-width Gaussian kernel is prohibitively expensive, as the width of the kernel scales with the number of pixels over which it must be evaluated at each point. The first convolution step, from $R=300\,000$ to $R=10\,000$, uses a single fixed-width Gaussian kernel applied to the whole spectrum, which is computationally cheap and reduces the pixel count by roughly 90\%. The subsequent, more expensive variable-width convolution to the DESI LSF is then performed on this coarser, resampled grid, substantially reducing the overall computing time without a meaningful loss of information, since the DESI resolution is itself always well below $R=10\,000$.

We computed a total of $\sim$370\,000 synthetic spectra with 30 varying physical and chemical labels: 6 physical parameters (\teff, \logg, \feh, \vmic, \vmac and \vsini) and 24 individual elemental abundances (Li, C, N, O, Na, Al, Mg, Si, Ti, K, Ca, Mn, Cr, Ni, S, V, Sc, Ba, La, Eu, Y, Zn, Zr, Cu). The parameter space was selected to ensure full coverage of FGK stars across all evolutionary stages, with \teff and \logg drawn from within the convex hull of a set of MIST stellar evolutionary tracks \citep{2016ApJS..222....8D,2016ApJ...823..102C} to ensure physically plausible combinations. The bounds for the effective temperature are 4000 and 8000\,K, with \logg between 0.0 and 5.0\,dex and [Fe/H] between $-2.5$ and $+1.0$\,dex. The values of $v_{\rm mic}$ and $v_{\rm mac}$ were calculated for each star from empirical relations as a function of \teff, \logg and [Fe/H] (see \cite{ispec2014}), while $v\sin i$ was drawn uniformly between 2 and 20\,km\,s$^{-1}$. Unlike LRPayne, where all elemental abundances [X/Fe] were randomly drawn from a common range of $-0.5$ to $+0.5$\,dex, here the range of each element was tailored individually to better reflect the diversity of abundance patterns expected in the DESI sample; solar abundances were taken from \citet{asplund2009}. All parameters except for \vmic and \vmac are drawn randomly and ensured to have a uniform distribution. The adopted ranges for all labels are summarised in Table~\ref{tab:label_ranges}. Note that, in the training, we do not use \vmac as one of the labels because at DESI's resolution, its effects are indistinguishable from \vsini.

Finally, to account for interstellar reddening, each synthetic spectrum was extincted following the \citet{2023ApJ...950...86G} extinction law, assuming $R_V=3.1$. A random value of $E(B-V)$ was drawn for each spectrum from a uniform distribution between 0 and 1\,mag and applied to the flux as $F_\lambda^{\rm ext} = F_\lambda \times 10^{-0.4\,A(\lambda)}$. $E(B-V)$ was then appended as an additional label, bringing the total number of labels used to train the network to 31. This allows the network to fit interstellar extinction simultaneously with the stellar parameters and abundances, rather than requiring the observed spectra to be de-reddened beforehand.

\begin{table}
\centering
\caption{Parameter space used in this study.}
\label{tab:label_ranges}
\begin{tabular}{lc}
\hline\hline
Label & Range \\
\hline
\teff & [4000, 8000] (K) \\
\logg & [0.0, 5.0] (dex) \\
{[Fe/H]} & [$-2.5$, $+1.0$] (dex) \\
$v_{\rm mic}$ & empirical (\teff, \logg, [Fe/H]) \\
$v_{\rm mac}$ & empirical (\teff, \logg, [Fe/H]) \\
$v\sin i$ & [2, 20] (km s$^{-1}$) \\
$E(B-V)$ & [0.0, 1.0] (mag) \\
\hline
{[C/Fe]}  & [$-0.5$, $+0.5$] \\
{[N/Fe]}  & [$-0.5$, $+1.0$] \\
{[O/Fe]}  & [$-0.5$, $+0.7$] \\
{[Na/Fe]} & [$-0.5$, $+0.6$] \\
{[Mg/Fe]} & [$-0.5$, $+0.8$] \\
{[Al/Fe]} & [$-0.5$, $+0.7$] \\
{[Si/Fe]} & [$-0.5$, $+0.8$] \\
{[K/Fe]}  & [$-0.5$, $+0.6$] \\
{[Ca/Fe]} & [$-0.5$, $+0.8$] \\
{[Sc/Fe]} & [$-0.5$, $+0.6$] \\
{[Ti/Fe]} & [$-0.5$, $+0.9$] \\
{[V/Fe]}  & [$-0.5$, $+0.5$] \\
{[Cr/Fe]} & [$-0.5$, $+0.5$] \\
{[Mn/Fe]} & [$-0.5$, $+0.5$] \\
{[Ni/Fe]} & [$-0.5$, $+0.5$] \\
{[Cu/Fe]} & [$-0.5$, $+0.5$] \\
{[Zn/Fe]} & [$-0.5$, $+0.5$] \\
{[Li/Fe]} & [$-1.5$, $+1.0$] \\
{[Ba/Fe]} & [$-1.5$, $+0.8$] \\
{[La/Fe]} & [$-0.8$, $+1.0$] \\
{[Eu/Fe]} & [$-0.5$, $+1.5$] \\
{[Y/Fe]}  & [$-0.8$, $+0.8$] \\
{[Zr/Fe]} & [$-0.4$, $+0.9$] \\
\hline
\end{tabular}
\end{table}

\subsection{Training}\label{sec:training}

The neural network is trained on a grid of $\sim$370,000 synthetic spectra described in Sect.\ref{sec:synthetic}. We split this grid into training and validation sets with an 80-20\% ratio. The network is optimised using the stochastic optimiser ADAM \citep{adam2014}. We use an initial learning rate of 0.001, together with a learning rate scheduler that reduces the learning rate by a factor of 0.5 whenever the validation loss has not improved for 10 epochs. For the loss function, we use the mean absolute percentage error. We found this to be more effective than a simple mean squared error at guiding the gradient descent. Training proceeds through alternating forward and backward propagation steps. Weights are first initialised using the He initialiser \citep{2015arXiv150201852H} for each of the three hidden layers. In the forward pass, the network produces a prediction from the current weights, which is then compared to the target flux to compute the loss. This loss is back-propagated to update the weights before the next forward pass. Throughout training, we use the validation set to monitor convergence and guard against overfitting. This provides an unbiased estimate of the network's performance on spectra it was not trained on.

Once trained, the network acts as a fast spectral interpolator. Given an arbitrary set of stellar labels, it returns the corresponding flux-calibrated SED, which we refer to as the constructed spectrum. To fit an observed DESI spectrum, we use the \texttt{curve\_fit} routine from the \texttt{scipy} package. This performs a $\chi^2$ minimisation between the constructed spectrum and the observed spectrum. We simultaneously fit for the 30 stellar labels together with the coefficients of the low-order polynomial that absorbs residual continuum mismatch caused by instrumental effects. We adopt a fit tolerance of $10^{-10}$, which we find balances reliable convergence against computational cost.

\subsection{DESI data}

DESI is a highly multiplexed fibre-fed spectrograph mounted on the Mayall 4-metre telescope at Kitt Peak National Observatory, capable of simultaneously observing 5000 targets over an 8 deg$^2$ field of view \citep{desi2016}. Each fiber spectrum is split by a dichroic beam splitter into three arms: B (3600-5930 \AA), R (5600-8000 \AA), and Z (7470-9800 \AA), giving continuous wavelength coverage from 3600 to 9800 \AA. The resolving power of the instrument varies with wavelength and arm, from $R\sim2000$ at the blue end to $R\sim5500$ at the red end \citep{koposov2025}. All DESI spectra are extracted at a fixed sampling of 0.8 \AA\ per pixel in the barycentric frame and are flux-calibrated using standard stars observed in each exposure \citep{guy2023}. In addition to its primary cosmological programme, DESI conducts the Milky Way Survey (MWS) in bright time, targeting stars down to $r\sim19$ \citep{cooper2023}. Stellar spectra are processed and classified by the standard DESI pipeline, before being passed to dedicated stellar pipelines (RVS and SP) that measure radial velocities, stellar parameters and abundances \citep{koposov2025}.

We use stellar spectra from the DESI Data Release 1 (DR1) Milky Way Survey Value-Added Catalogue \citep{koposov2025}, specifically the merged catalogue \texttt{mwsall-pix-iron.fits}. It combines RVS and SP pipeline measurements across all surveys and programmes (main, sv1-3, and special), along with the crossmatch to \textit{Gaia} DR3. For sources observed in more than one survey or programme, this catalogue retains only the highest signal-to-noise (S/N, in R band) observation as the primary measurement, yielding 5,930,124 unique objects, of which 4,378,110 are classified as stars. We take the co-added, flux-calibrated spectra corresponding to these primary observations directly from the DESI data release, without any of the continuum normalisation applied by the official RVS or SP pipelines. 

\subsection{Masking}

Fitting observed spectra with a neural network trained on synthetic models is complicated by the fact that the network cannot perfectly reproduce every observed spectrum, due to the approximate treatment of stellar atmospheric physics (e.g. NLTE and 3D effects), incompleteness of the linelist, and inaccuracies in atomic data. Strong, saturated lines are particularly affected, as their cores form high in the atmosphere where these approximations break down most severely, and small mismatches between the line profiles can dominate the $\chi^2$ and bias the fit. We therefore mask a number of wavelength regions before fitting. We mask the cores of the hydrogen Balmer lines (H$\alpha$, H$\beta$, H$\gamma$, H$\delta$), whose broad, pressure-sensitive wings are poorly reproduced by 1D LTE models. We also mask the Na~D doublet (5890, 5896~\AA), which is frequently affected by interstellar absorption in addition to the stellar photospheric contribution, and whose non-LTE core formation is not well captured by our LTE synthetic models. Similarly, we mask the Ca~II H \& K lines (3933, 3968~\AA), which are strong, saturated features whose cores form high in the chromosphere and are poorly reproduced by our 1D LTE models. In total, we masked 170 pixels or 2.5\% of the wavelength region. We also explored an alternative, threshold-based approach to identify additional bad pixels, in which we compared a high signal-to-noise solar spectrum, degraded to DESI resolution, with the corresponding synthetic spectrum computed at solar parameters, and masked any pixel where the flux deviation between the two exceeded a given threshold. However, this approach did not improve the results compared to masking only the strong lines described above, and we therefore adopt the simpler strong-line masking scheme for the rest of this work.

\section{Results}
\subsection{Internal Accuracy}
\begin{figure*}
    \centering
    \begin{subfigure}{0.99\linewidth}
        \centering
        \includegraphics[width=\linewidth]{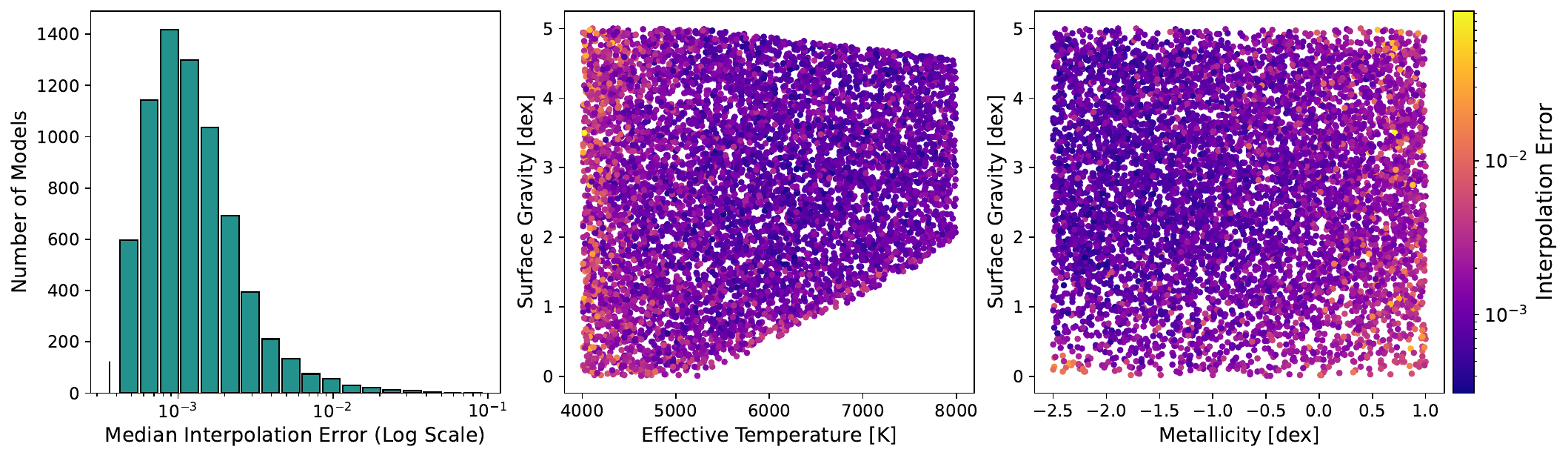}
        \caption{Histogram of median interpolation errors from the ANN for 7000 synthetic models (left); correlation of interpolation errors with effective temperature and surface gravity of the synthetic models (middle); and correlation of interpolation error with metallicity and surface gravity (right) of the synthetic models.}
        \label{fig:interpolation}
    \end{subfigure}

    \vspace{1em}

    \begin{subfigure}{0.85\linewidth}
        \centering
        \includegraphics[width=\linewidth]{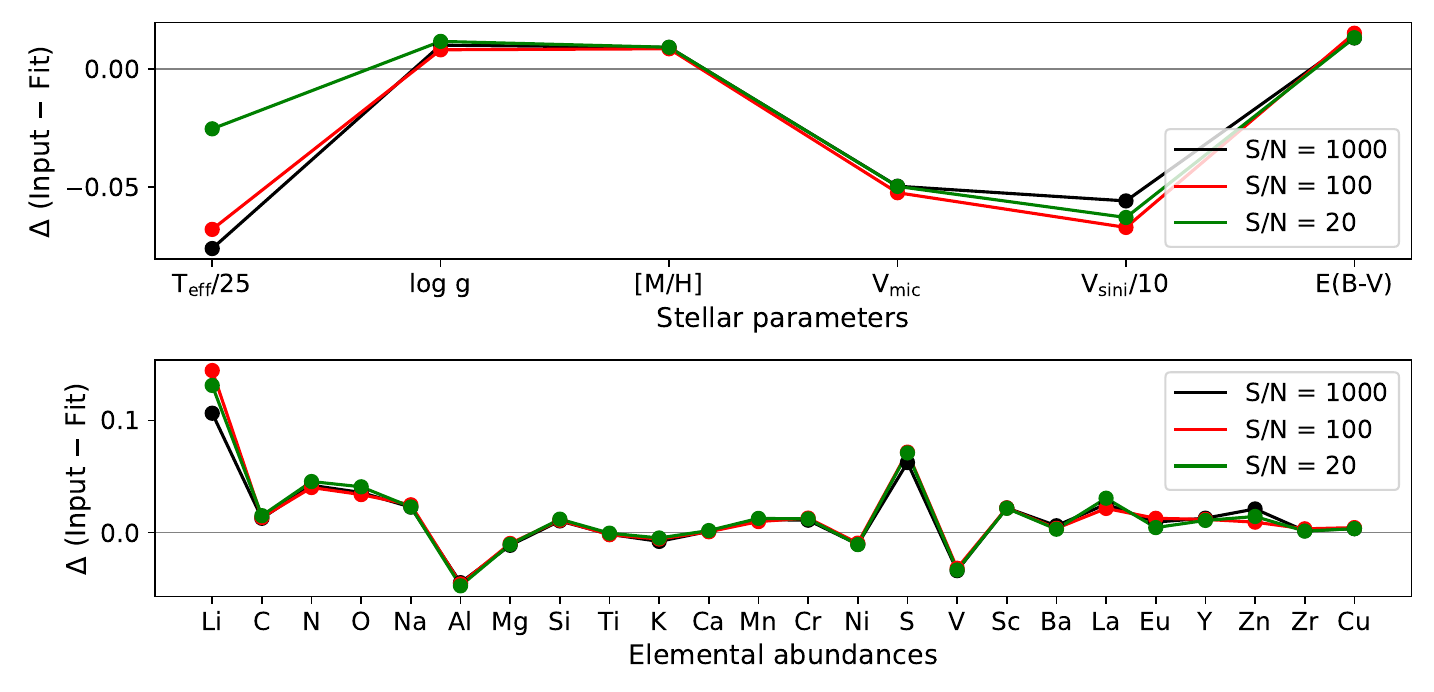}
        \caption{Comparison of input labels with values recovered by the neural network from fitting multiple synthetic spectra at different S/N (green: S/N = 20, red: S/N = 100 and black: S/N = 1000).}
        \label{fig:syn_verification}
    \end{subfigure}

    \caption{}
    \label{fig:interp_and_verification}
\end{figure*}

To verify the performance of the trained network, we use two internal accuracy tests, following the same approach as \citet{Payne}. Both tests make use of an additional set of 7000 synthetic spectra, generated in the same way as the training grid (see Section \ref{sec:synthetic}) but withheld from training, which we refer to as the synthetic verification sample. The first test examines the interpolation accuracy of the neural network. For each spectrum in the verification sample, we compare the flux generated by Turbospectrum with the flux interpolated by the network for the same input labels. Figure~\ref{fig:interpolation} shows the results of this test. The left panel shows the distribution of the median interpolation error across the verification sample. The distribution peaks around a median error of $\sim10^{-3}$, with the large majority of the sample below $10^{-2}$ and only a small tail of models reaching errors as large as $10^{-1}$. The middle and right panels show how the interpolation error correlates with \teff, \logg, and [Fe/H]. The largest interpolation errors are found preferentially near the edges of the parameter space: in particular, for cool, high-\logg\ dwarfs at the low-\teff\ boundary of the grid, and for stars at the metal-poor and metal-rich edges of the metallicity range. This is expected, as models near the boundary of the training grid have fewer neighbouring synthetic spectra for the network to interpolate between, and cool, metal-rich stars in particular have spectra that are rich in blended lines and molecular bands, making them intrinsically harder to interpolate accurately. Overall, this result demonstrates that the interpolation accuracy of the neural network is high across the great majority of the parameter space explored in this work (90\,$\%$ of the spectra have interpolation error of less than 0.003 or 0.3 $\%$). 

The second test examined how well the entire workflow could recover labels from synthetic models using $\chi^{2}$ minimisation. For this, we used the same synthetic validation sample at three different S/N ratios: 1000, 100, and 20. We used $add\_noise$ function within iSpec to degrade the S/N of the synthetic spectra. Although the added noise did not fully represent the real noise in an observed spectrum, it was sufficient to establish an upper limit on the accuracy of the model to recover the labels at different S/N. Figure~\ref{fig:syn_verification} shows the median differences between the input and recovered values for the 30 stellar parameters and abundances. For the stellar parameters, we find good agreement between input and recovered values across all three S/N. \logg, \feh, and $E(B-V)$ show small, consistent offsets of order 0.02 that do not depend strongly on S/N. In \teff, we find better performance at S/N = 20 than at S/N = 1000; however, the difference is less than a kelvin. This could be due to \teff being dominated by the SED shape and not the spectral lines. $v_{\rm mic}$ and $v\sin i$ show larger median offsets of order 0.05 and 0.5, respectively, again with little dependence on S/N. For the elemental abundances, the majority of elements are recovered with a median offset close to zero at all three S/N, and, unlike the trend seen for \teff, the recovery of most labels shows little degradation between S/N = 1000 and S/N = 20. The largest discrepancy is found for Li (0.15 dex) and S (0.08 dex), which is expected given the lack of detectable, unblended lines at DESI resolution. 

\subsection{Verification on observed spectra}\label{sec:obs_veri}

Upon satisfactory results from the internal accuracy tests, the network needs to be tested for its ability to reliably recover stellar labels from real observed spectra. To do this, we cross-matched the DESI catalogue with APOGEE DR17 \citep{2022ApJS..259...35A} and GALAH DR3 \citep{2021MNRAS.506..150B} to find stars in common. We required APOGEE stars to satisfy \texttt{FE\_H\_FLAG}~=~0 and \texttt{STARFLAG}~=~0, and GALAH stars to satisfy \texttt{flag\_sp}~=~0 and \texttt{flag\_fe\_h}~=~0, following the quality cuts recommended by \citet{koposov2025}. We additionally required a median DESI signal-to-noise ratio of at least 20 for all stars in the comparison. Of the cross-matched sample, 6719 APOGEE and 3455 GALAH stars satisfied these quality and S/N cuts. We fitted all of these stars using our trained network and compared the resulting labels with those from the two surveys. The results are shown in Figure~\ref{fig:parameter_initial_guess}. Within the fitting, we used \teff, \logg and \feh from \citet{2025ApJ...984...58H} as an initial guess, derived from Gaia DR3 XP spectra. This initial guess is available for 1,378,325 of the 1,727,374 stars in DESI with S/N $\ge$ 20, or 80\% of the sample. For the remaining 20\% that lack a Gaia XP-based initial guess, we use a separate fitting mechanism, described and verified in Section~\ref{sec:global_fitting}. This is necessary because the network fits the SED, extinction, and 30 individual labels simultaneously. The resulting degeneracies between labels mean that, without a good initial guess, the standard fitting procedure can easily become trapped in a local minimum and return unreliable labels. We estimate that an initial guess that is off by more than 300 K in \teff and 0.3 dex in \logg and \feh can lead to unreliable labels. We note that the coverage of the Gaia XP initial guess increases sharply with S/N, from 54\% at S/N~=~10 to 90\% at S/N~=~50. All results from Section \ref{sec:parameters} to \ref{sec:S/N dependency} only discuss stars with initial guesses from Gaia XP.

\subsubsection{Parameters} \label{sec:parameters}
\begin{figure*}
    \centering
    \includegraphics[width=0.99\linewidth]{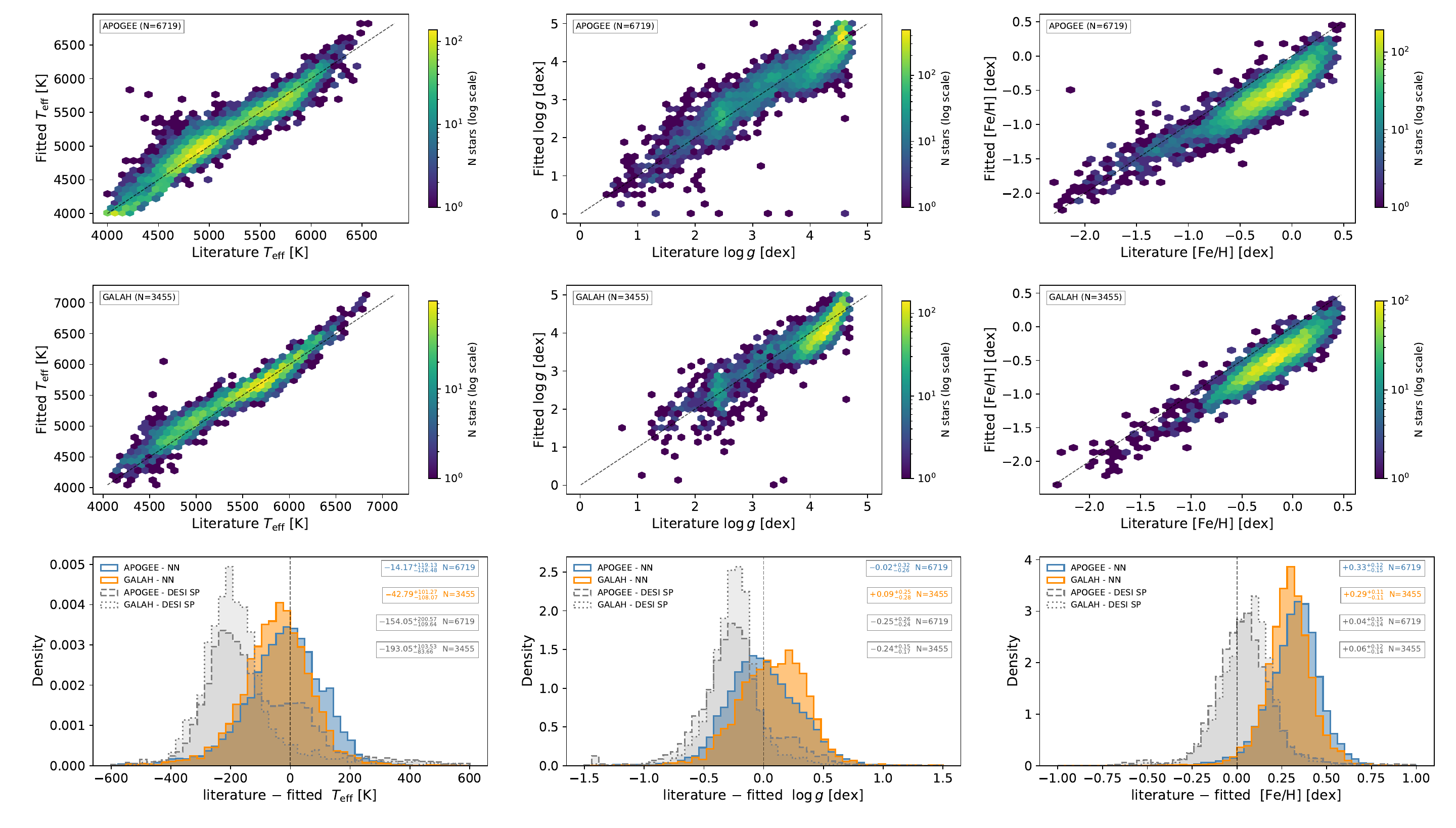}
    \caption{Comparison of parameters obtained from the neural network with those from APOGEE and GALAH. In the bottom row, grey-dashed and grey-dotted correspond to residuals between DESI's official SP model with that of APOGEE and GALAH, respectively. Metallicity from DESI SP has already been corrected. }
    \label{fig:parameter_initial_guess}
\end{figure*}

Figure~\ref{fig:parameter_initial_guess} compares stellar parameters recovered by our network with literature values from APOGEE (top row) and GALAH (middle row). The bottom row shows the corresponding distributions of residuals, together with the equivalent residuals for the DESI SP pipeline (FERRE) on the same stars, for comparison. 
For \teff, our network shows good agreement with both external surveys, with median offsets of $-14^{+119}_{-126}$~K against APOGEE and $-47^{+105}_{-110}$~K against GALAH. This represents a substantial improvement over the DESI SP pipeline, which shows systematic offsets of $-154^{+201}_{-110}$~K and $-193^{+108}_{-85}$~K against the same APOGEE and GALAH samples, respectively. We attribute this improvement largely to our decision to fit the flux-calibrated SED directly, rather than a continuum-normalised spectrum as done by SP: the overall shape of the SED is itself a sensitive tracer of \teff, and this information is discarded once the spectrum is normalised, leaving only the line-profile shapes to constrain the temperature. By retaining the SED shape, our network is able to constrain \teff using both the continuum shape and the line profiles simultaneously, which likely explains the reduced scatter and systematic offset relative to SP. For \logg, our network again performs well, with median offsets close to zero ($-0.02^{+0.32}_{-0.26}$~dex against APOGEE, $+0.09^{+0.25}_{-0.28}$~dex against GALAH), compared to a systematic offset of $\sim-0.25$~dex found for SP against both surveys. Figure \ref{fig:HRdiagram} shows the Hertzsprung–Russell (HR) diagram for the GALAH + APOGEE sample using results from the neural network fitting. The main sequence, subgiant branch, and red giant branch are all well-populated and cleanly separated, confirming that our network recovers physically sensible stellar parameters across the full range of evolutionary stages present in the DESI MWS sample. For \feh, however, our network shows a systematic offset of $+0.33^{+0.12}_{-0.15}$~dex against APOGEE and $+0.29^{+0.11}_{-0.11}$~dex against GALAH, indicating that our fitted metallicities are on average lower than the literature values by $\sim$0.3~dex. When compared to uncorrected \feh from DESI SP, our offsets are larger and in the opposite direction, with DESI SP showing an offset of $-0.07^{+0.24}_{-0.10}$ dex in APOGEE and $+0.05^{+0.12}_{-0.13}$ in GALAH. However, the APOGEE distribution in DESI SP is bimodal, which is not seen in our result. The bimodality is not visible in Figure \ref{fig:parameter_initial_guess} because the figure uses corrected metallicities from DESI SP (see Sect. 4.2 in \cite{koposov2025}. We attribute our systematic underestimation of \feh\ to an improper estimation of the continuum zero point by the low-order polynomial used to absorb residual continuum mismatch. At the moderate S/N typical of DESI spectra, we find that this polynomial tends to over-estimate the zero point of the continuum while still correctly recovering its overall shape. This leads to an effective over-subtraction of the continuum level, which in turn biases the recovered metallicity toward lower values.

\begin{figure}
    \centering
    \includegraphics[width=0.99\linewidth]{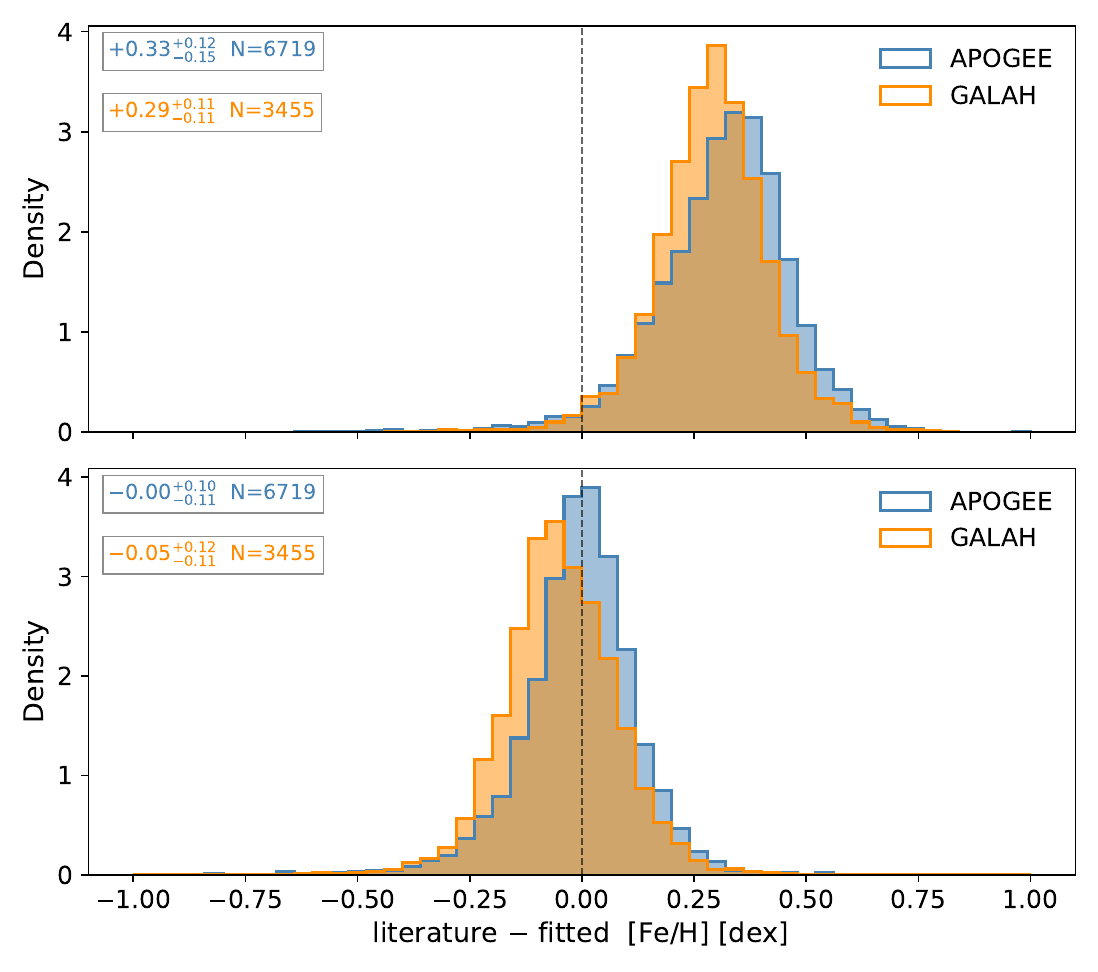}
    \caption{Comparison of uncorrected (top) and corrected (bottom) metallicity distributions for APOGEE (blue) and GALAH (orange).}
    \label{fig:feh_corrected}
\end{figure}

\begin{figure}
    \centering
    \includegraphics[width=0.99\linewidth]{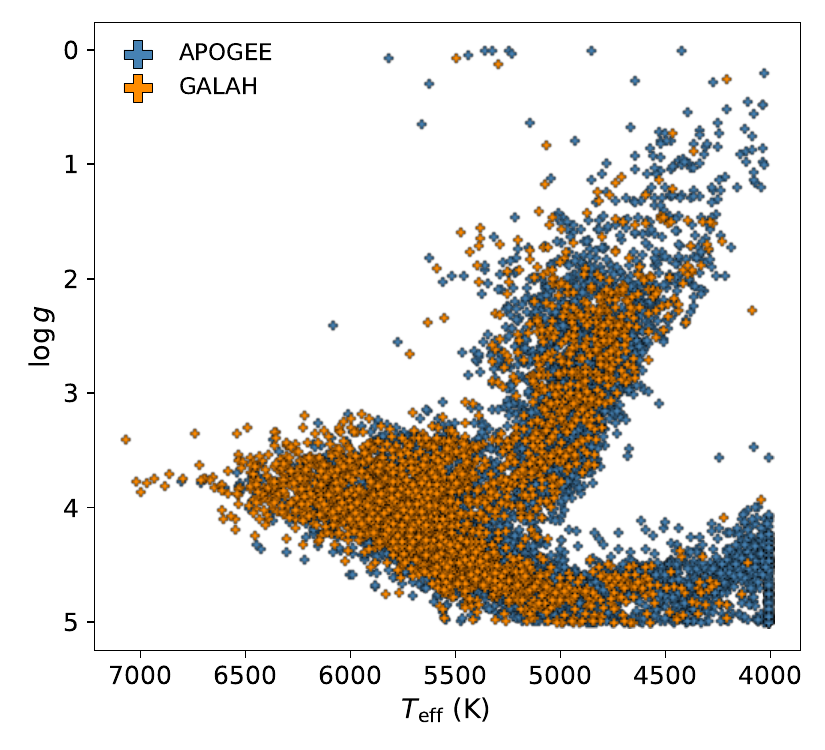}
    \caption{Hertzsprung–Russell (HR) diagram of the APOGEE (blue) and GALAH (orange) cross-match sample fitted using our neural network.}
    \label{fig:HRdiagram}
\end{figure}

To mitigate this offset, we applied the same temperature- and surface-gravity-dependent calibration described in Sect.~4.2.1 of \citet{koposov2025}, in which a quadratic function of effective temperature, with separate coefficients for dwarfs and giants, is subtracted from the raw [Fe/H] to correct for this systematic trend. We refit $a$, $b$, and $c$ coefficients separately for dwarfs and giants (given in Table \ref{tab:feh_calib}) using our APOGEE cross-match sample, following the same optimisation procedure as \citet{koposov2025}. Since the raw [Fe/H] offsets against APOGEE and GALAH are very similar (Fig.~\ref{fig:feh_corrected}, top panel), we apply the same APOGEE-derived coefficients to correct both samples, rather than fitting GALAH separately. Figure~\ref{fig:feh_corrected} shows the distribution of residuals before and after applying this calibration. After calibration, the offsets are reduced to $-0.00^{+0.10}_{-0.11}$~dex and $-0.05^{+0.12}_{-0.12}$~dex, respectively, bringing our [Fe/H] into agreement with SP at a comparable level of accuracy. We publish both the uncorrected and calibrated [Fe/H] values in our catalogue, allowing users to apply an alternative calibration if desired, or to work directly with the raw metallicities if preferred. The coefficients of the calibration fit are given in \ref{tab:feh_calib}.

\begin{table}
\centering
\caption{Coefficients of the quadratic [Fe/H] calibration, fit to the APOGEE cross-match sample and applied to both APOGEE and GALAH comparisons.}
\label{tab:feh_calib}
\begin{tabular}{lccccc}
\hline\hline
Subset & $N$ & $a$ & $b$ & $c$  \\
\hline
Giants ($\log g < 3.7$) & 2418 & $+0.2220$ & $+0.0241$ & $+0.1937$ \\
Dwarfs ($\log g \geq 3.7$) & 4301 & $+0.3511$ & $-0.0255$ & $+0.0796$  \\
\hline
\end{tabular}
\end{table}

We note that the SP residuals shown here are larger than those reported for the same comparison samples in \citet{koposov2025}. This is because the official SP catalogue additionally excludes stars fitted with the secondary PHOENIX grid (\texttt{BESTGRID}~$\neq$~\texttt{'s\_rdesi1'}), a cut that we do not apply here since it is not relevant to our method. This highlights one advantage of our approach: because the network is trained on a single, continuous grid of synthetic spectra, all stars are fitted homogeneously.





\subsection{Abundances}\label{Sec:abundance}

Even though the neural network is trained to fit 24 individual abundances, in this section we only discuss 12 of them: Na, Mg, Al, Si, Ca, Ti, V, Cr, Mn, Ni, Ba, and Y. The reason for this is the high accuracy of the neural network for these elements at DESI resolution. For other elements such as Li, C, N, O, K, and S, the network shows significantly worse accuracy, and these will not be included in the final catalogue.

\begin{figure*}
    \centering
    \includegraphics[width=0.99\linewidth]{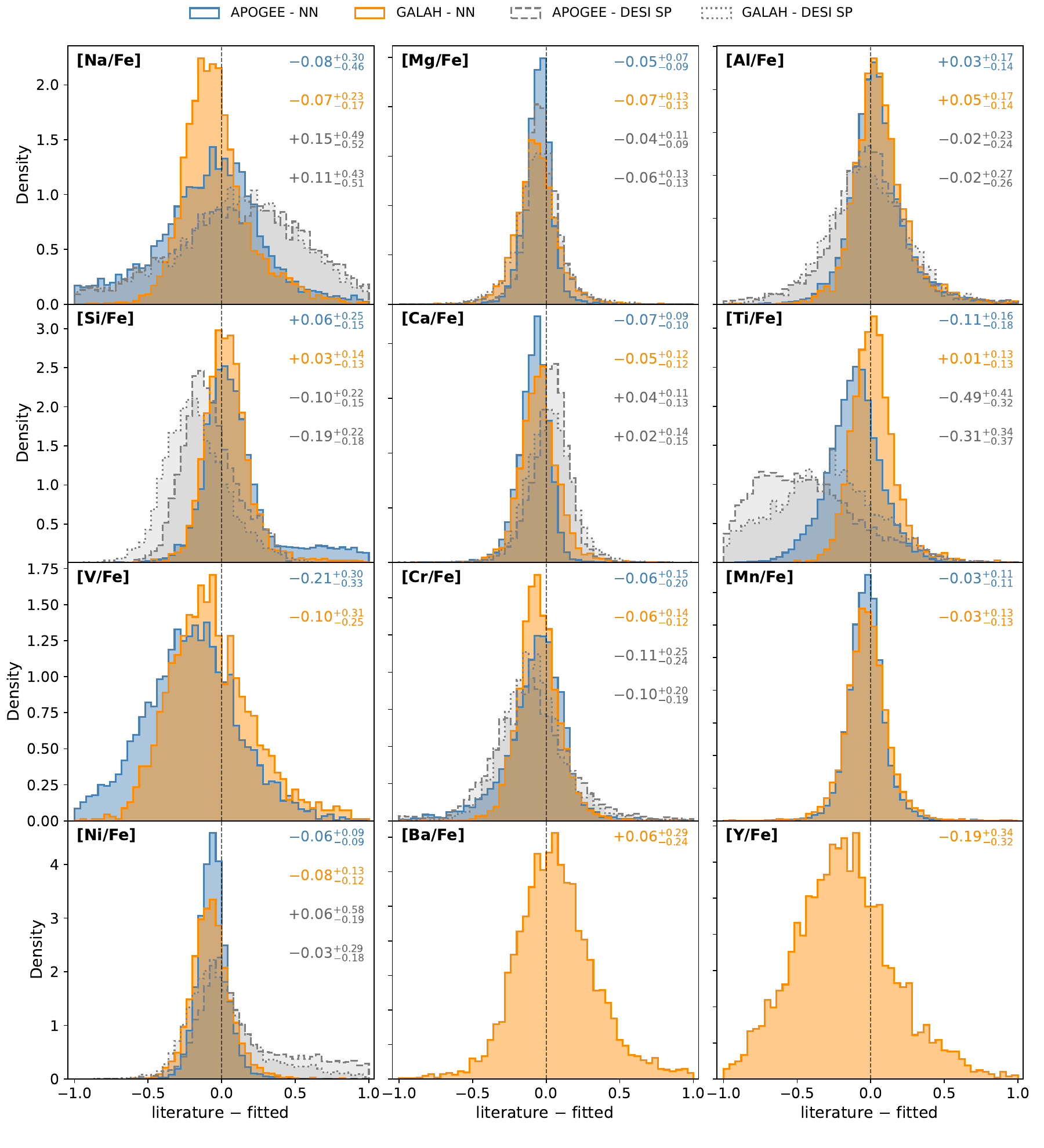}
    \caption{Comparison of abundances obtained from the neural network with those from APOGEE and GALAH. The grey-dashed and grey-dotted histograms correspond to residuals between DESI's official SP model with that of APOGEE and GALAH, respectively. Of the 24 elements, 12 of them plotted here are estimated well.  }
    \label{fig:abundance_initial_guess}
\end{figure*}

Figure~\ref{fig:abundance_initial_guess} compares the abundances recovered by our network with literature values from APOGEE and GALAH, following the same cross-match samples and quality cuts as used for the stellar parameters (Sect.~\ref{sec:parameters}). SP reports abundances as [X/H] rather than [X/Fe]; to allow a direct comparison, we converted these to [X/Fe] using the calibrated SP metallicities. We also show the equivalent residuals for the DESI SP pipeline, where available. Beyond providing more reliable abundances for several of the elements SP does measure, our method also delivers abundances for four elements not provided by SP at all: Mn, V, Ba, and Y.

\subsubsection{$\alpha$-elements}

\begin{figure}
    \centering
    \includegraphics[width=0.99\linewidth]{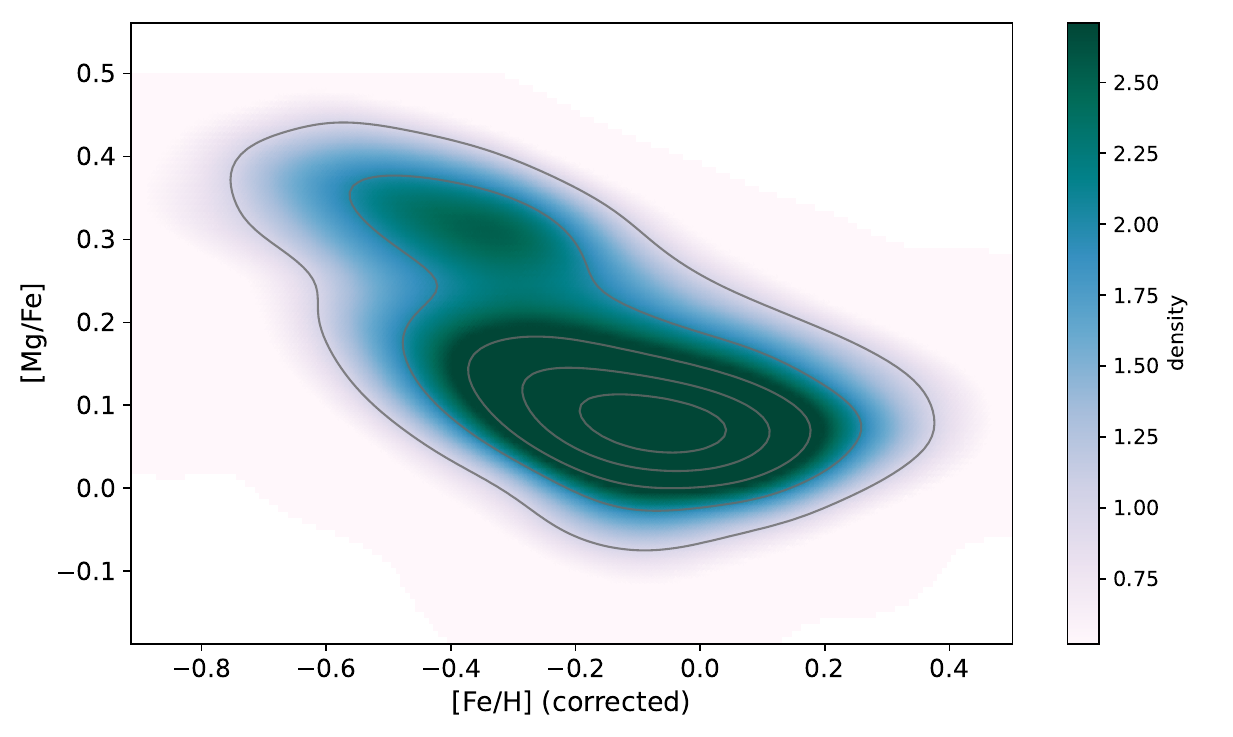}
    \caption{Distribution of [Mg/Fe] versus corrected [Fe/H] for the APOGEE cross-match sample. The chemical bimodality of the Milky Way disc, the high-[$\alpha$/Fe] thick disc and low-[$\alpha$/Fe] thin disc, is clearly recovered.}
    \label{fig:MgH}
\end{figure}

Among the $\alpha$-elements, we verify four abundances: Mg, Si, Ca, and Ti. Our network recovers [Mg/Fe] and [Ca/Fe] with small offsets of $-0.05$ to $-0.07$~dex against both APOGEE and GALAH, comparable to SP ($-0.04$ to $-0.06$~dex for Mg, $+0.02$ to $+0.04$~dex for Ca). For [Si/Fe], our offsets are smaller ($+0.03$ to $+0.06$~dex) than SP's ($-0.10$ to $-0.19$~dex). In Figure \ref{fig:MgH}, we examine the distribution of [Mg/Fe] with respect to [Fe/H] for the APOGEE sample. The two sequences of the Milky Way disc — the high-[Mg/Fe] thick disc and the low-[Mg/Fe] thin disc — are clearly recovered, consistent with the well-established bimodal chemical structure seen in higher-resolution surveys \citep{apogee2017, 2021MNRAS.506..150B}. 

For [Ti/Fe], our network shows an offset of $0.11$~dex against APOGEE, but only $0.01$~dex against GALAH. Ti abundances from APOGEE are known to be unreliable \citep{2015AJ....150..148H}, and the excellent agreement with GALAH is therefore a more meaningful measure of our network's performance for this element. By comparison, SP shows large, systematic offsets of $-0.49$~dex against APOGEE and $-0.31$~dex against GALAH, together with a substantially broader and visibly bimodal residual distribution (Fig.~\ref{fig:abundance_initial_guess}). This is consistent with \citet{koposov2025}, who note that only Mg, Ca, and Fe abundances from SP are considered reliable. This result shows that our network is able to recover Ti, and by extension Si, with a higher level of accuracy than the SP pipeline does. 

\subsubsection{Odd-Z elements}

Within the odd-Z elements, we verify Na and Al. For [Na/Fe], our network shows an offset of $-0.08$~dex against APOGEE and $+0.07$~dex against GALAH, with a somewhat asymmetric residual distribution reflecting the difficulty of measuring Na at low resolution; this nonetheless represents an improvement over SP, whose offsets of $+0.15$ and $+0.11$~dex are larger and show a broader scatter, with a pronounced tail toward positive residuals. For [Al/Fe], our network shows small offsets of $+0.03$ to $+0.05$~dex, again comparable to or better than SP ($-0.02$~dex for both surveys, but with larger scatter of 0.23-0.27~dex compared to our 0.14-0.17~dex).

\subsubsection{Iron-peak elements}

Within the iron-peak elements, we verify V, Cr, Mn, and Ni. [Cr/Fe] and [Ni/Fe] are recovered with small offsets ($-0.06$ to $-0.08$~dex) and modest scatter for both surveys. For [Ni/Fe] against APOGEE, SP shows a markedly broader and asymmetric residual distribution ($+0.06^{+0.58}_{-0.19}$~dex). [Mn/Fe] shows excellent agreement with both surveys, with offsets of only $-0.03$~dex and scatter of $\sim0.11$-$0.13$~dex. [V/Fe] shows the largest offset among the iron-peak elements, at $-0.21$~dex against APOGEE and $-0.10$~dex against GALAH, with correspondingly broader scatter. Vanadium abundances from APOGEE are also known to be unreliable \citep{2020AJ....160..120J}, similarly to Ti, so the smaller offset against GALAH is again the more representative measure of our network's performance. SP does not report both Mn and V.

\subsubsection{Heavy elements}

Among the heavy elements, we verify two s-process elements, Ba and Y, using the GALAH comparison sample only, as neither element is available in our APOGEE cross-match. [Ba/Fe] is recovered with a small offset of $+0.06$~dex, though with substantial scatter ($^{+0.29}_{-0.24}$~dex). [Y/Fe] shows a larger offset of $-0.19$~dex and similarly broad scatter ($^{+0.34}_{-0.32}$~dex). Neither element is reported by SP. While the scatter for both elements is broad, reflecting the intrinsic difficulty of measuring weak neutron-capture lines at DESI's resolution and signal-to-noise, the ability to recover these abundances at all for a good fraction of stars is valuable in its own right: applied to the full DESI sample, this would provide heavy-element abundances for hundreds of thousands of stars.

\subsubsection{Dependency on S/N}\label{sec:S/N dependency}

\begin{figure}
    \centering
    \includegraphics[width=0.99\linewidth]{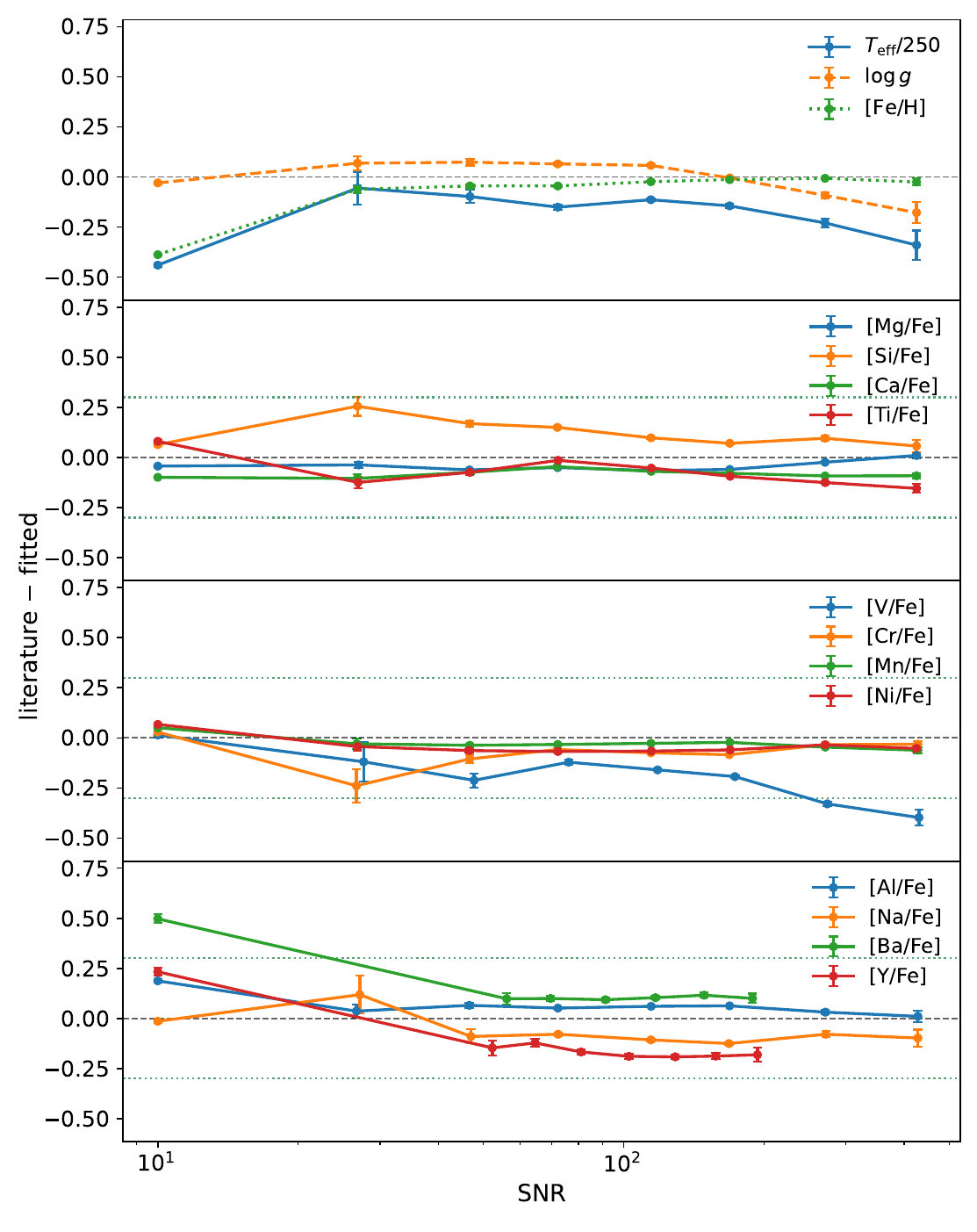}
    \caption{Median residual as a function of DESI S/N for the stellar parameters (top panel) and the 12 abundances (remaining panels), computed from the combined APOGEE and GALAH cross-match sample. The central points of the S/N bins are at 10, 25, 50, 75, 110, 170, 270, and 400. Dotted horizontal lines mark $\pm0.3$~dex, corresponding to the satisfactory-accuracy threshold adopted throughout this work.}
    \label{fig:SNR_dependency}
\end{figure}

Since DESI stars span a wide range in S/N, it is important to understand how the accuracy of our fitted labels depends on it. This is particularly important for DESI, where roughly half of the $\sim$6 million stellar spectra have $\mathrm{S/N}>10$, with the number dropping to 1.7 million if we restrict to $\mathrm{S/N}\geq20$. To maximise the scientific return of our catalogue, it is therefore essential that our method can reliably recover stellar labels even at S/N as low as 10. To test this, we bin our combined APOGEE and GALAH comparison sample by S/N and examine the median residual for each label as a function of S/N. Figure~\ref{fig:SNR_dependency} shows the results for the stellar parameters and for the 12 abundances. The central points of the SNR bins are at 10, 25, 50, 75, 110, 170, 270 and 400. In the original APOGEE + GALAH comparison sample, no spectra have S/N less than 20. Therefore, for the results at S/N~$=10$, we degraded the S/N of the sample to 10 and refit them.

For the stellar parameters, the performance at medium to high S/N is good in all three. But at S/N~$=10$, the accuracy of \teff\ and \feh\ gets worse compared to \logg, which has almost a 0.0~dex offset. At high S/N, we see the accuracy of \teff\ and \logg\ get worse compared to \feh. For the abundances, most elements show flat or only mildly varying offsets across the full S/N range, including at S/N~$=10$: [Mg/Fe], [Ca/Fe], [Mn/Fe], [Ni/Fe], and [Na/Fe] all remain close to their typical offset. [Al/Fe] and [Y/Fe] are a bit higher at S/N~$=10$ ($\sim+0.19$ and $\sim+0.24$~dex, respectively) but stay within the typical offset range. [Ba/Fe] is the exception: it shows the largest offset of any abundance at low S/N, $\sim+0.50$~dex at S/N~$=10$, above our offset threshold, and drops steadily to $\sim+0.1$~dex by S/N~$\sim 75$--$110$. [Si/Fe] and [Cr/Fe] both show small offsets at S/N~$=10$ that increase to a peak at S/N~$\sim 25$, before coming back down to near zero by S/N~$\sim 100$. [V/Fe] is near zero at S/N~$=10$, becomes mildly negative ($\sim-0.1$ to $-0.2$~dex) through the mid-S/N range, and grows more negative to $\sim-0.4$~dex at the highest S/N bin, the same pattern we see for \teff\ and \logg\ at high S/N. Overall, these results show that most labels are recovered reliably down to S/N~$=10$.

\subsection{Fitting Without an Initial Guess}\label{sec:global_fitting}

For the 40\% of the S/N$\,\geq 10$ DESI sample (drops to 20\% for S/N$\,\geq 20$) that lack a Gaia XP-based initial guess (see Sect.~\ref{sec:obs_veri}), the standard fitting procedure cannot be used directly. Without a good starting point in the 30-dimensional label space, the $\chi^2$ minimisation readily converges to a local minimum, owing to the degeneracies between labels discussed above, and returns unreliable stellar parameters and abundances.

To fit these stars, we instead adopt a two-step, grid-based approach designed to avoid local minima without relying on an external initial guess. In the first step, we perform a coarse grid search over the four labels most strongly affected by degeneracies: \teff, \logg, \feh, and $E(B-V)$. We use a fixed grid of $8\times5\times7\times6\approx1700$ points spanning the full range of the training grid in these four labels, while all remaining labels ($v_{\rm mic}$, $v\sin i$, and all abundances) are held fixed at the centre of their scaled range. At each grid point, the four free labels, together with the fixed remaining labels, are passed through the network to generate a constructed spectrum. Since the polynomial continuum term enters the model linearly, its coefficients can be solved in closed form via weighted least squares at each grid point, without requiring a non-linear search. We evaluate the $\chi^2$ of the resulting constructed spectrum against the observed spectrum at every grid point, and adopt the point with the lowest $\chi^2$ as the starting solution. Because this comparison is made across the full grid rather than by following a local gradient from a single seed, it is able to correctly disentangle \teff from $E(B-V)$: a gradient-based method started on the wrong side of this degeneracy would instead converge to whichever local minimum it started closer to.

In the second step, we expand the best-fitting grid point back into the full 30-dimensional label vector and refine it using a bounded local optimiser (\texttt{scipy.optimize.curve\_fit} with the trust-region-reflective method). Every label is therefore still fit without any external information; the grid search in the first step simply identifies the correct minimum more robustly than a single non-linear optimisation from an uninformed starting point would. We also tested differential evolution as an alternative global optimiser, but we found it to be extremely computation-heavy and would not be possible for fitting a couple of hundred thousand stars. We refer to this two-step procedure as the ``global'' fit, in contrast to the standard fit that uses the Gaia XP initial guess directly.

\begin{figure}
    \centering
    \includegraphics[width=0.99\linewidth]{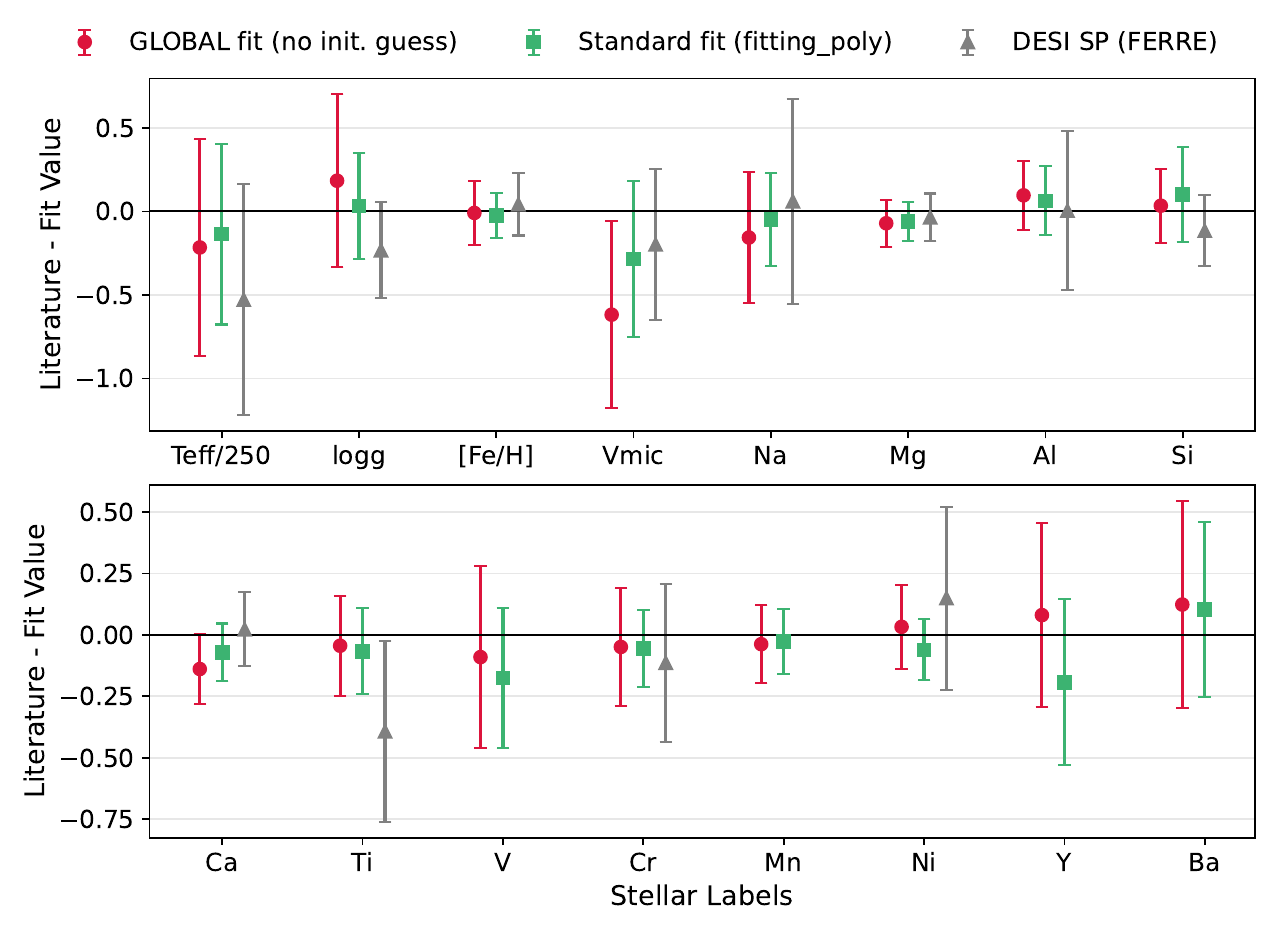}
    \caption{Comparison of labels obtained from global fitting without initial guesses (\textit{red}), standard fitting with initial guesses (\textit{green}) and the DESI SP pipeline with the labels from APOGEE and GALAH. The point and the error bar represent the mean and standard deviation of the residual distributions. }
    \label{fig:global_fitting}
\end{figure}

To validate this approach, we apply it to the same APOGEE and GALAH cross-match sample used in Sect.~\ref{sec:obs_veri}, deliberately withholding the Gaia XP initial guess, and compare the recovered labels to both the standard, initial-guess-based fit and to the literature values. Figure~\ref{fig:global_fitting} shows the resulting mean and standard deviation for the stellar parameters and abundances, together with the equivalent DESI SP values for reference. Global fitting performs comparably to the standard fitting for almost all the labels. It reduces the offsets in \teff, \logg, Si, Ti and Ni compared to the DESI SP pipeline, but is accompanied by an increase in the standard deviation (in the case of \teff and \logg). The standard deviation in abundances such as Na, Al, Ti and Ni, though slightly larger than standard fitting, is still an improvement over the DESI SP pipeline. This shows that even for objects without initial guesses, we will be able to provide reliable stellar parameters and abundances.

\subsection{Uncertainties}

Our fitting workflow is subject to two distinct sources of parameter uncertainty. The first is the fitting error arising from the $\chi^2$ minimisation. The second is the model-to-model variation introduced by the random weight initialisation used during ANN training. To capture both simultaneously, we adopted an MCMC-like sampling scheme. We trained ten separate ANN models on different random shufflings of the training set, with all hyperparameters held fixed. Each star was then processed through the full workflow 1000 times per model, varying only the initial guess supplied to the $\chi^2$ minimiser. Repeating this across all ten models yielded 10,000 individual fits per star. The resulting posterior distribution for each parameter was used to define the $1\sigma$ uncertainty as the interval between the 16th and 84th percentiles. In this way, the repeated $\chi^2$ minimisation captures the fitting uncertainty, while the ensemble of independently trained models captures the uncertainty due to network initialisation. 

We estimated uncertainties for seven stars spanning different regions of the parameter space, at three different S/N (10, 20, and 50). Table~\ref{tab:err_all} gives the resulting uncertainties. We do not provide an estimate for a Hot Rich Giant, since no star in the APOGEE/GALAH comparison sample with $\mathrm{S/N}\geq120$ satisfies all of the required conditions for that combination of \teff and \logg.

For most stars, the stellar parameters have low uncertainties compared to traditional methods. The one outlier is a Hot Metal-Poor Giant, which has an uncertainty of $\sim$71~K in \teff and $\sim$0.26~dex in \logg at the lowest S/N probed. Across the 12 abundances, we find excellent accuracy for at least six elements: Mg, Si, Ca, Ti, Cr, and Mn, where the uncertainty stays below 0.2~dex at every S/N, with the exception of the Hot Poor Giant. For elements such as Na, Al, Si, and V, we find large errors in hot stars but smaller errors in cool stars. This is because these elements have very few good lines in the optical, which become very weak in the spectra of hot stars. In these cases, noise dominates the line signal, leading to larger errors.


\begin{deluxetable}{llcccccccc}
\tabletypesize{\scriptsize}
\tablecaption{Uncertainties on Derived Stellar Parameters and Abundance Ratios as a Function of S/N\label{tab:err_all}}
\tablehead{
\colhead{Star} & \colhead{S/N} & \colhead{$\sigma_{T_{\rm eff}}$ (K)} & \colhead{$\sigma_{\log g}$} & \colhead{$\sigma_{\rm [Fe/H]}$} & \colhead{$\sigma_{V_{\rm mic}}$} & \colhead{$\sigma_{\rm Na/Fe}$} & \colhead{$\sigma_{\rm Mg/Fe}$} & \colhead{$\sigma_{\rm Al/Fe}$} & \colhead{$\sigma_{\rm Si/Fe}$} \\
}
\startdata
\multirow{3}{*}{\textbf{Hot Metal-Rich Dwarf}} & 10 & 75 & 0.16 & 0.07 & 0.21 & 0.20 & 0.08 & \textbf{0.50} & 0.08 \\
 & 20 & 40 & 0.11 & 0.06 & 0.21 & 0.16 & 0.07 & 0.12 & 0.07 \\
 & 50 & 30 & 0.08 & 0.03 & 0.10 & 0.09 & 0.04 & 0.16 & 0.04 \\
\tableline
\multirow{3}{*}{\textbf{Hot Metal-Poor Dwarf}} & 10 & 60 & 0.17 & 0.07 & 0.20 & \textbf{0.84} & 0.13 & \textbf{0.46} & 0.08 \\
 & 20 & 50 & 0.11 & 0.07 & 0.19 & \textbf{0.31} & 0.11 & 0.11 & 0.08 \\
 & 50 & 20 & 0.07 & 0.03 & 0.12 & 0.16 & 0.05 & 0.09 & 0.05 \\
\tableline
\multirow{3}{*}{\textbf{Hot Metal-Poor Giant}} & 10 & 70 & 0.26 & 0.13 & \textbf{0.34} & \textbf{0.86} & \textbf{0.33} & \textbf{0.66} & \textbf{0.70} \\
 & 20 & 40 & 0.20 & 0.10 & 0.20 & \textbf{0.55} & 0.18 & \textbf{0.42} & \textbf{0.42} \\
 & 50 & 40 & 0.15 & 0.08 & 0.23 & \textbf{0.63} & 0.19 & 0.24 & 0.28 \\
\tableline
\multirow{3}{*}{\textbf{Cool Metal-Rich Dwarf}} & 10 & 70 & 0.11 & 0.08 & 0.24 & 0.18 & 0.05 & 0.08 & 0.19 \\
 & 20 & 50 & 0.10 & 0.07 & 0.24 & 0.25 & 0.04 & 0.07 & 0.07 \\
 & 50 & 50 & 0.09 & 0.06 & 0.21 & 0.15 & 0.03 & 0.09 & 0.07 \\
\tableline
\multirow{3}{*}{\textbf{Cool Metal-Rich Giant}} & 10 & 60 & 0.19 & 0.07 & 0.28 & \textbf{0.49} & 0.04 & 0.12 & 0.08 \\
 & 20 & 60 & 0.14 & 0.07 & 0.29 & \textbf{0.41} & 0.04 & 0.13 & 0.06 \\
 & 50 & 60 & 0.14 & 0.06 & 0.27 & \textbf{0.40} & 0.04 & 0.11 & 0.05 \\
\tableline
\multirow{3}{*}{\textbf{Cool Metal-Poor Dwarf}} & 10 & 54 & 0.12 & 0.06 & 0.24 & 0.11 & 0.04 & 0.09 & 0.12 \\
 & 20 & 50 & 0.09 & 0.05 & 0.19 & 0.12 & 0.03 & 0.09 & 0.08 \\
 & 50 & 50 & 0.09 & 0.04 & 0.14 & 0.10 & 0.03 & 0.06 & 0.07 \\
\tableline
\multirow{3}{*}{\textbf{Cool Metal-Poor Giant}} & 10 & 77 & 0.17 & 0.07 & \textbf{0.33} & 0.16 & 0.05 & 0.23 & 0.07 \\
 & 20 & 50 & 0.14 & 0.04 & 0.11 & 0.12 & 0.03 & 0.08 & 0.05 \\
 & 50 & 40 & 0.15 & 0.04 & 0.12 & 0.12 & 0.03 & 0.09 & 0.04 \\
\tableline
\textbf{Star} & \textbf{S/N} & \textbf{$\sigma_{\rm Ca/Fe}$} & \textbf{$\sigma_{\rm Ti/Fe}$} & \textbf{$\sigma_{\rm V/Fe}$} & \textbf{$\sigma_{\rm Cr/Fe}$} & \textbf{$\sigma_{\rm Mn/Fe}$} & \textbf{$\sigma_{\rm Ni/Fe}$} & \textbf{$\sigma_{\rm Y/Fe}$} & \textbf{$\sigma_{\rm Ba/Fe}$} \\
\tableline
\multirow{3}{*}{\textbf{Hot Metal-Rich Dwarf}} & 10 & 0.05 & 0.08 & 0.11 & 0.09 & 0.16 & 0.11 & 0.23 & 0.20 \\
 & 20 & 0.04 & 0.06 & 0.14 & 0.12 & 0.08 & 0.06 & 0.16 & 0.23 \\
 & 50 & 0.02 & 0.04 & 0.07 & 0.03 & 0.04 & 0.03 & 0.15 & 0.23 \\
\tableline
\multirow{3}{*}{\textbf{Hot Metal-Poor Dwarf}} & 10 & 0.06 & 0.13 & 0.20 & 0.06 & 0.12 & 0.21 & \textbf{0.41} & \textbf{0.40} \\
 & 20 & 0.05 & 0.06 & 0.18 & 0.05 & 0.07 & 0.11 & 0.25 & \textbf{0.31} \\
 & 50 & 0.03 & 0.03 & 0.11 & 0.05 & 0.05 & 0.05 & 0.10 & 0.26 \\
\tableline
\multirow{3}{*}{\textbf{Hot Metal-Poor Giant}} & 10 & \textbf{0.31} & 0.18 & 0.11 & \textbf{0.34} & \textbf{0.35} & 0.16 & 0.15 & \textbf{0.52} \\
 & 20 & 0.15 & 0.13 & \textbf{0.57} & 0.20 & 0.23 & \textbf{0.46} & \textbf{0.35} & \textbf{0.72} \\
 & 50 & 0.11 & 0.07 & \textbf{0.46} & 0.16 & 0.10 & 0.23 & 0.27 & \textbf{0.55} \\
\tableline
\multirow{3}{*}{\textbf{Cool Metal-Rich Dwarf}} & 10 & 0.04 & 0.07 & 0.10 & 0.05 & 0.06 & 0.06 & 0.17 & \textbf{0.43} \\
 & 20 & 0.03 & 0.05 & 0.13 & 0.04 & 0.06 & 0.04 & 0.15 & 0.15 \\
 & 50 & 0.03 & 0.05 & 0.10 & 0.04 & 0.04 & 0.04 & 0.12 & 0.15 \\
\tableline
\multirow{3}{*}{\textbf{Cool Metal-Rich Giant}} & 10 & 0.06 & 0.10 & 0.25 & 0.07 & 0.09 & 0.07 & 0.29 & 0.18 \\
 & 20 & 0.05 & 0.07 & 0.08 & 0.04 & 0.07 & 0.05 & 0.17 & 0.14 \\
 & 50 & 0.05 & 0.06 & 0.10 & 0.04 & 0.06 & 0.04 & 0.10 & 0.12 \\
\tableline
\multirow{3}{*}{\textbf{Cool Metal-Poor Dwarf}} & 10 & 0.04 & 0.06 & \textbf{0.32} & 0.05 & 0.05 & 0.14 & \textbf{0.59} & \textbf{0.34} \\
 & 20 & 0.03 & 0.04 & 0.11 & 0.04 & 0.04 & 0.04 & 0.23 & 0.16 \\
 & 50 & 0.03 & 0.03 & 0.07 & 0.03 & 0.04 & 0.04 & 0.29 & 0.08 \\
\tableline
\multirow{3}{*}{\textbf{Cool Metal-Poor Giant}} & 10 & 0.05 & 0.07 & 0.10 & 0.07 & 0.13 & 0.07 & 0.15 & \textbf{0.61} \\
 & 20 & 0.04 & 0.04 & 0.06 & 0.04 & 0.09 & 0.05 & 0.18 & 0.11 \\
 & 50 & 0.04 & 0.04 & 0.06 & 0.03 & 0.10 & 0.03 & 0.18 & 0.14 \\
\enddata
\vspace{1em}
\resizebox{0.8\textwidth}{!}{%
\begin{tabular}{llll}
\textbf{Hot Metal-Rich Dwarf:} 6186\,K, 4.14\,dex, 0.05\,dex &
\textbf{Hot Metal-Poor Dwarf:} 6135\,K, 4.30\,dex, -0.54\,dex &
\textbf{Hot Metal-Poor Giant:} 6264\,K, 3.76\,dex, -0.60\,dex &\\
\textbf{Cool Metal-Rich Dwarf:} 4990\,K, 4.42\,dex, -0.17\,dex &
\textbf{Cool Metal-Rich Giant:} 4748\,K, 2.29\,dex, -0.07\,dex &
\textbf{Cool Metal-Poor Dwarf:} 5329\,K, 4.38\,dex, -0.55\,dex &\\
\textbf{Cool Metal-Poor Giant:} 4754\,K, 2.58\,dex, -0.71\,dex &
\end{tabular}%
}
\end{deluxetable}

\subsection{Catalogue} \label{catalog}

We will release our results as a machine-readable FITS catalogue containing stellar parameters, abundances, and their associated uncertainties for DESI DR1 spectra with $\mathrm{S/N}\geq10$\footnote{For reference to the referee, we have published a partial catalogue with 170,000 stars at \url{https://doi.org/10.5281/zenodo.21816591}}. Stars with a Gaia XP-based initial guess are fit using the standard procedure; stars without a Gaia XP guess are fit using the global grid-search procedure. The \texttt{HAS\_GAIA\_GUESS} column records which of the two procedures was used for each star. The catalogue contains 3.2 million rows in total.

To flag potentially unreliable fits, we compute a quality flag following \citet{zhang2024},
\begin{equation}
    \mathrm{qflag}_{\chi^2} = \frac{\chi^2 - \chi^2_{\rm median}}{\chi^2_{\rm mad}},
\end{equation}
where $\chi^2_{\rm median}$ and $\chi^2_{\rm mad}$ are the median and mean absolute deviation of $\chi^2$ computed within bins of width 5 in S/N, up to $\mathrm{S/N}=150$ (all higher-S/N stars are grouped into the final bin), using only bins containing at least 20 stars. This flag identifies stars whose fit is anomalously poor relative to other stars of similar spectral quality, rather than using an absolute $\chi^2$ threshold, which would otherwise disfavour certain spectra due to the nature of the data. We recommend excluding stars with \texttt{CHI2\_FLAG} $>$ 3 from science samples requiring the most reliable fits. Table~\ref{tab:catalogue} describes the columns included in the catalogue.

\begin{table}
\centering
\small
\setlength{\tabcolsep}{4pt}
\caption{Description of the columns provided in the catalogue. }
\label{tab:catalogue}
\begin{tabular}{ll}
\hline\hline
Field & Description \\
\hline
TARGETID & DESI target identifier \\
GAIA\_SOURCE\_ID & \textit{Gaia} DR3 source ID ($-1$ if no match) \\
HAS\_GAIA\_GUESS & True: standard fit; False: global fit \\
RA & Right ascension (deg) \\
DEC & Declination (deg) \\
SNR\_MED & Median DESI spectral S/N \\
RV & DESI pipeline radial velocity (km~s$^{-1}$) \\
RV\_ERR & Uncertainty in RV (km~s$^{-1}$) \\
TEFF & Effective temperature (K) \\
TEFF\_ERR & Fitting error in TEFF (K) \\
LOGG & Surface gravity (dex) \\
LOGG\_ERR & Fitting error in LOGG (dex) \\
FEH\_RAW & Uncalibrated [Fe/H] (dex) \\
FEH & Calibrated [Fe/H]  \\
FEH\_ERR & Fitting error in FEH (dex) \\
VMIC & Microturbulent velocity (km~s$^{-1}$) \\
VMIC\_ERR & Fitting error in VMIC (km~s$^{-1}$) \\
CHI2 & $\chi^2$ of the best fit \\
CHI2\_FLAG & S/N-relative $\chi^2$ quality flag \\
X\_FE & [X/Fe] for element X (dex) \\
X\_FE\_ERR & Uncertainty in X\_FE (dex) \\
\hline
\end{tabular}
\end{table}

\section{Conclusion}

We have presented a machine-learning pipeline for determining stellar parameters and chemical abundances from DESI spectra, adapting the model-driven approach of The Payne to the flux-calibrated, un-normalised SEDs delivered by DESI. Rather than fitting a continuum-normalised spectrum, as is standard practice for large spectroscopic surveys, we train our network to reproduce the observed SED directly, including interstellar extinction as a fitted label alongside the stellar parameters and 24 elemental abundances. Our main findings are summarised below.

\begin{itemize}
\item \textbf{Architecture and internal accuracy.} A five-layer network with three hidden layers of 2000 neurons each, using a LeakyReLU activation throughout, including the output layer, provides accurate interpolation across our full training grid of $\sim$370,000 synthetic spectra. Internal accuracy tests show a median interpolation error below 0.3\% for 90\% of the verification sample, and the full fitting workflow recovers input labels reliably even at $\mathrm{S/N}=20$.

\item \textbf{Stellar parameters.} Validated against 6719 APOGEE and 3455 GALAH stars observed by DESI, our network recovers $T_{\rm eff}$ and $\log g$ with smaller offsets than the official DESI SP pipeline, which we attribute to our use of the full, un-normalised SED shape rather than a continuum-normalised spectrum. The raw [Fe/H] shows a systematic offset of $\sim$0.3~dex, likely a side effect of the low-order polynomial used to absorb residual continuum mismatch; applying a temperature- and gravity-dependent calibration, following the same approach used for the official DESI pipelines, reduces this offset to zero. We publish both the raw and calibrated [Fe/H] in our catalogue.

\item \textbf{Abundances.} Of the 24 elements our network is trained to fit, we identify 12, Na, Mg, Al, Si, Ca, Ti, V, Cr, Mn, Ni, Ba, and Y, that are recovered with sufficient accuracy to include in our final catalogue. For elements also measured by SP, our results are comparable or substantially better, most notably for Ti and Si. We additionally provide Mn, V, Ba, and Y, none of which are reported by the official DESI pipelines, extending the chemical information available for DESI stars beyond what is currently possible with existing pipelines.

\item \textbf{Robustness to signal-to-noise.} Most of our fitted labels show little to no degradation in accuracy across the S/N range covered by our comparison sample ($\mathrm{S/N}\gtrsim20$), indicating that our method is not strongly limited by spectral noise for the majority of stars. This robustness is essential given that a large fraction of the DESI stellar sample is observed at moderate S/N.

\item \textbf{Fitting without an initial guess.} For stars lacking a Gaia XP-based initial guess, roughly 40\% of our $\mathrm{S/N}\geq10$ sample, we developed a two-step grid-search-and-polish fitting strategy that avoids the local minima that otherwise plague the standard fit. We show that this ``global'' fitting approach recovers labels of comparable quality to the standard method, allowing us to provide reliable parameters and abundances across the full sample regardless of external initial-guess availability.

\end{itemize}

Applying this pipeline to the DESI Data Release 1 stellar sample, we deliver a catalogue of stellar parameters and abundances for 3.2 million stars, substantially expanding the sample of DESI stars with homogeneously determined chemical abundances beyond what is available from the official DESI pipelines. This catalogue will support studies of Galactic chemical evolution, stellar population membership, and the identification of chemically peculiar or metal-poor stars across the large and diverse stellar sample delivered by DESI. Future DESI data releases, expected to roughly triple the size of the current stellar sample, will allow this method to be applied to an even larger and more complete census of the Galaxy. The catalogue and trained model will be publicly available before the article is published.

\begin{acknowledgments}

\end{acknowledgments}
This work presents results from the European Space Agency (ESA) space mission Gaia. Gaia data are being processed by the Gaia Data Processing and Analysis Consortium (DPAC). Funding for the DPAC is provided by national institutions, in particular the institutions participating in the Gaia MultiLateral Agreement (MLA). The Gaia mission website is https://www.cosmos.esa.int/gaia. The Gaia archive website is https://archives.esac.esa.int/gaia.

For our analysis, we made use of \texttt{astropy} \citep{astropy2013, astropy2018, astropy2022}, \texttt{scipy} \citep{virtanen2020}, \texttt{numpy} \citep{harris2020}, \texttt{pandas} \citep{pandas}, and \texttt{matplotlib} \citep{hunter2007} for general data handling, fitting, and visualisation. Our neural network was built and trained using \texttt{TensorFlow} \citep{abadi2015}. Synthetic spectra were generated with \texttt{iSpec} \citep{ispec2014, ispec2019} and the Turbospectrum synthesis code \citep{turbospectrum}, and interstellar extinction was applied using \texttt{dust\_extinction} \citep{dust_extinction}. Spectral data were handled in HDF5 format using \texttt{h5py} \citep{h5py}. We also made use of TOPCAT \citep{taylor2005} for catalogue cross-matching and inspection.

\section{Data Availability}

The full catalogue will be provided as a FITS file uploaded to Zenodo as well as in VIZIER once the full sample is fitted. We expect this to be done before the final publication of the paper. The training code and the trained model will be made available on GitHub at the same time. For reference of the referee, we have published the partial catalogue for 170,000 stars in Zenodo at \url{https://doi.org/10.5281/zenodo.21816591}.

\begin{contribution}

All authors contributed equally to the publication.


\end{contribution}

\bibliography{sample701}{}

@ARTICLE{Asplund2009,
       author = {{Asplund}, Martin and {Grevesse}, Nicolas and {Sauval}, A. Jacques and {Scott}, Pat},
        title = "{The Chemical Composition of the Sun}",
      journal = {\araa},
         year = 2009,
        month = sep,
       volume = {47},
       number = {1},
        pages = {481-522},
          doi = {10.1146/annurev.astro.46.060407.145222},
archivePrefix = {arXiv},
       eprint = {0909.0948},
 primaryClass = {astro-ph.SR},
       adsurl = {https://ui.adsabs.harvard.edu/abs/2009ARA&A..47..481A}
}

@ARTICLE{2002ARA&A..40..487F,
       author = {{Freeman}, Ken and {Bland-Hawthorn}, Joss},
        title = "{The New Galaxy: Signatures of Its Formation}",
      journal = {\araa},
         year = 2002,
        month = jan,
       volume = {40},
        pages = {487-537},
          doi = {10.1146/annurev.astro.40.060401.093840},
archivePrefix = {arXiv},
       eprint = {astro-ph/0208106},
 primaryClass = {astro-ph},
       adsurl = {https://ui.adsabs.harvard.edu/abs/2002ARA&A..40..487F}
}

@ARTICLE{2016ApJ...823..102C,
       author = {{Choi}, Jieun and {Dotter}, Aaron and {Conroy}, Charlie and {Cantiello}, Matteo and {Paxton}, Bill and {Johnson}, Benjamin D.},
        title = "{Mesa Isochrones and Stellar Tracks (MIST). I. Solar-scaled Models}",
      journal = {\apj},
         year = 2016,
        month = jun,
       volume = {823},
       number = {2},
          eid = {102},
        pages = {102},
          doi = {10.3847/0004-637X/823/2/102},
archivePrefix = {arXiv},
       eprint = {1604.08592},
 primaryClass = {astro-ph.SR},
       adsurl = {https://ui.adsabs.harvard.edu/abs/2016ApJ...823..102C}
}

@ARTICLE{2016ApJS..222....8D,
       author = {{Dotter}, Aaron},
        title = "{MESA Isochrones and Stellar Tracks (MIST) 0: Methods for the Construction of Stellar Isochrones}",
      journal = {\apjs},
         year = 2016,
        month = jan,
       volume = {222},
       number = {1},
          eid = {8},
        pages = {8},
          doi = {10.3847/0067-0049/222/1/8},
archivePrefix = {arXiv},
       eprint = {1601.05144},
 primaryClass = {astro-ph.SR},
       adsurl = {https://ui.adsabs.harvard.edu/abs/2016ApJS..222....8D}
}

@ARTICLE{2020AJ....160..120J,
       author = {{J{\"o}nsson}, Henrik and {Holtzman}, Jon A. and {Allende Prieto}, Carlos and {Cunha}, Katia and {Garc{\'\i}a-Hern{\'a}ndez}, D.~A. and {Hasselquist}, Sten and {Masseron}, Thomas and {Osorio}, Yeisson and {Shetrone}, Matthew and {Smith}, Verne and {Stringfellow}, Guy S. and {Bizyaev}, Dmitry and {Edvardsson}, Bengt and {Majewski}, Steven R. and {M{\'e}sz{\'a}ros}, Szabolcs and {Souto}, Diogo and {Zamora}, Olga and {Beaton}, Rachael L. and {Bovy}, Jo and {Donor}, John and {Pinsonneault}, Marc H. and {Poovelil}, Vijith Jacob and {Sobeck}, Jennifer},
        title = "{APOGEE Data and Spectral Analysis from SDSS Data Release 16: Seven Years of Observations Including First Results from APOGEE-South}",
      journal = {\aj},
         year = 2020,
        month = sep,
       volume = {160},
       number = {3},
          eid = {120},
        pages = {120},
          doi = {10.3847/1538-3881/aba592},
archivePrefix = {arXiv},
       eprint = {2007.05537},
 primaryClass = {astro-ph.GA},
       adsurl = {https://ui.adsabs.harvard.edu/abs/2020AJ....160..120J}
}

@ARTICLE{2015AJ....150..148H,
       author = {{Holtzman}, Jon A. and {Shetrone}, Matthew and {Johnson}, Jennifer A. and {Allende Prieto}, Carlos and {Anders}, Friedrich and {Andrews}, Brett and {Beers}, Timothy C. and {Bizyaev}, Dmitry and {Blanton}, Michael R. and {Bovy}, Jo and {Carrera}, Ricardo and {Chojnowski}, S. Drew and {Cunha}, Katia and {Eisenstein}, Daniel J. and {Feuillet}, Diane and {Frinchaboy}, Peter M. and {Galbraith-Frew}, Jessica and {Garc{\'\i}a P{\'e}rez}, Ana E. and {Garc{\'\i}a-Hern{\'a}ndez}, D.~A. and {Hasselquist}, Sten and {Hayden}, Michael R. and {Hearty}, Fred R. and {Ivans}, Inese and {Majewski}, Steven R. and {Martell}, Sarah and {Meszaros}, Szabolcs and {Muna}, Demitri and {Nidever}, David and {Nguyen}, Duy Cuong and {O'Connell}, Robert W. and {Pan}, Kaike and {Pinsonneault}, Marc and {Robin}, Annie C. and {Schiavon}, Ricardo P. and {Shane}, Neville and {Sobeck}, Jennifer and {Smith}, Verne V. and {Troup}, Nicholas and {Weinberg}, David H. and {Wilson}, John C. and {Wood-Vasey}, W.~M. and {Zamora}, Olga and {Zasowski}, Gail},
        title = "{Abundances, Stellar Parameters, and Spectra from the SDSS-III/APOGEE Survey}",
      journal = {\aj},
         year = 2015,
        month = nov,
       volume = {150},
       number = {5},
          eid = {148},
        pages = {148},
          doi = {10.1088/0004-6256/150/5/148},
archivePrefix = {arXiv},
       eprint = {1501.04110},
 primaryClass = {astro-ph.GA},
       adsurl = {https://ui.adsabs.harvard.edu/abs/2015AJ....150..148H}
}

@ARTICLE{2021MNRAS.506..150B,
       author = {{Buder}, Sven and {Sharma}, Sanjib and {Kos}, Janez and {Amarsi}, Anish M. and {Nordlander}, Thomas and {Lind}, Karin and {Martell}, Sarah L. and {Asplund}, Martin and {Bland-Hawthorn}, Joss and {Casey}, Andrew R. and {de Silva}, Gayandhi M. and {D'Orazi}, Valentina and {Freeman}, Ken C. and {Hayden}, Michael R. and {Lewis}, Geraint F. and {Lin}, Jane and {Schlesinger}, Katharine J. and {Simpson}, Jeffrey D. and {Stello}, Dennis and {Zucker}, Daniel B. and {Zwitter}, Toma{\v{z}} and {Beeson}, Kevin L. and {Buck}, Tobias and {Casagrande}, Luca and {Clark}, Jake T. and {{\v{C}}otar}, Klemen and {da Costa}, Gary S. and {de Grijs}, Richard and {Feuillet}, Diane and {Horner}, Jonathan and {Kafle}, Prajwal R. and {Khanna}, Shourya and {Kobayashi}, Chiaki and {Liu}, Fan and {Montet}, Benjamin T. and {Nandakumar}, Govind and {Nataf}, David M. and {Ness}, Melissa K. and {Spina}, Lorenzo and {Tepper-Garc{\'\i}a}, Thor and {Ting}, Yuan-Sen and {Traven}, Gregor and {Vogrin{\v{c}}i{\v{c}}}, Rok and {Wittenmyer}, Robert A. and {Wyse}, Rosemary F.~G. and {{\v{Z}}erjal}, Maru{\v{s}}a and {Galah Collaboration}},
        title = "{The GALAH+ survey: Third data release}",
      journal = {\mnras},
         year = 2021,
        month = sep,
       volume = {506},
       number = {1},
        pages = {150-201},
          doi = {10.1093/mnras/stab1242},
archivePrefix = {arXiv},
       eprint = {2011.02505},
 primaryClass = {astro-ph.GA},
       adsurl = {https://ui.adsabs.harvard.edu/abs/2021MNRAS.506..150B}
}

@ARTICLE{ispec2014,
       author = {{Blanco-Cuaresma}, S. and {Soubiran}, C. and {Heiter}, U. and {Jofr{\'e}}, P.},
        title = "{Determining stellar atmospheric parameters and chemical abundances of FGK stars with iSpec}",
      journal = {\aap},
         year = 2014,
        month = sep,
       volume = {569},
          eid = {A111},
        pages = {A111},
          doi = {10.1051/0004-6361/201423945},
archivePrefix = {arXiv},
       eprint = {1407.2608},
 primaryClass = {astro-ph.IM},
       adsurl = {https://ui.adsabs.harvard.edu/abs/2014A&A...569A.111B}
}

@MISC{abadi2015,
       author = {{Abadi}, M. and {Agarwal}, A. and {Barham}, P. and et al.},
        title = "{TensorFlow: Large-Scale Machine Learning on Heterogeneous Systems}",
         year = 2015,
          url = {https://www.tensorflow.org/}
}

@ARTICLE{astropy2013,
       author = {{Astropy Collaboration} and {Robitaille}, T.~P. and {Tollerud}, E.~J. and et al.},
        title = "{Astropy: A community Python package for astronomy}",
      journal = {\aap},
         year = 2013,
       volume = {558},
          eid = {A33},
          doi = {10.1051/0004-6361/201322068}
}

@ARTICLE{astropy2018,
       author = {{Astropy Collaboration} and {Price-Whelan}, A.~M. and et al.},
        title = "{The Astropy Project: Building an Open-science Project and Status of the v2.0 Core Package}",
      journal = {\aj},
         year = 2018,
       volume = {156},
          eid = {123},
          doi = {10.3847/1538-3881/aabc4f}
}

@ARTICLE{astropy2022,
       author = {{Astropy Collaboration} and {Price-Whelan}, A.~M. and et al.},
        title = "{The Astropy Project: Sustaining and Growing a Community-oriented Open-source Project}",
      journal = {\apj},
         year = 2022,
       volume = {935},
          eid = {167},
          doi = {10.3847/1538-4357/ac7c74}
}

@ARTICLE{virtanen2020,
       author = {{Virtanen}, P. and {Gommers}, R. and {Oliphant}, T.~E. and et al.},
        title = "{SciPy 1.0: fundamental algorithms for scientific computing in Python}",
      journal = {Nature Methods},
         year = 2020,
       volume = {17},
        pages = {261},
          doi = {10.1038/s41592-019-0686-2}
}

@ARTICLE{harris2020,
       author = {{Harris}, C.~R. and {Millman}, K.~J. and {van der Walt}, S.~J. and et al.},
        title = "{Array programming with NumPy}",
      journal = {Nature},
         year = 2020,
       volume = {585},
        pages = {357},
          doi = {10.1038/s41586-020-2649-2}
}

@ARTICLE{hunter2007,
       author = {{Hunter}, J.~D.},
        title = "{Matplotlib: A 2D Graphics Environment}",
      journal = {Computing in Science and Engineering},
         year = 2007,
       volume = {9},
        pages = {90},
          doi = {10.1109/MCSE.2007.55}
}

@MISC{pandas,
       author = {{The pandas development team}},
        title = "{pandas-dev/pandas: Pandas}",
         year = 2020,
          doi = {10.5281/zenodo.3509134}
}

@MISC{h5py,
       author = {{Collette}, A. and et al.},
        title = "{h5py}",
         year = 2023,
          doi = {10.5281/zenodo.7568214}
}

@INPROCEEDINGS{taylor2005,
       author = {{Taylor}, M.~B.},
        title = "{TOPCAT \& STIL: Starlink Table/VOTable Processing Software}",
    booktitle = {Astronomical Data Analysis Software and Systems XIV},
         year = 2005,
       editor = {{Shopbell}, P. and {Britton}, M. and {Ebert}, R.},
       series = {ASP Conf. Ser.},
       volume = {347},
        pages = {29},
      address = {San Francisco, CA},
 organization = {ASP}
}

@MISC{dust_extinction,
       author = {{Gordon}, K.~D.},
        title = "{dust\_extinction}",
         year = 2024,
          url = {https://dust-extinction.readthedocs.io/}
}

@ARTICLE{apogee2017,
       author = {{Majewski}, S.~R. and {Schiavon}, R.~P. and {Frinchaboy}, P.~M. and {Allende Prieto}, C. and et al.},
        title = "{The Apache Point Observatory Galactic Evolution Experiment (APOGEE)}",
      journal = {\aj},
         year = 2017,
       volume = {154},
          eid = {94},
        pages = {94},
          doi = {10.3847/1538-3881/aa784d}
}

@ARTICLE{gaiaeso2012,
       author = {{Gilmore}, G. and {Randich}, S. and {Asplund}, M. and et al.},
        title = "{The Gaia-ESO Public Spectroscopic Survey}",
      journal = {The Messenger},
         year = 2012,
       volume = {147},
        pages = {25}
}

@ARTICLE{gaiaeso2013,
       author = {{Randich}, S. and {Gilmore}, G. and {Gaia-ESO Consortium}},
        title = "{The Gaia-ESO Large Public Spectroscopic Survey}",
      journal = {The Messenger},
         year = 2013,
       volume = {154},
        pages = {47}
}

@ARTICLE{galah2015,
       author = {{De Silva}, G.~M. and {Freeman}, K.~C. and {Bland-Hawthorn}, J. and et al.},
        title = "{The GALAH survey: scientific motivation}",
      journal = {\mnras},
         year = 2015,
       volume = {449},
        pages = {2604},
          doi = {10.1093/mnras/stv327}
}

@ARTICLE{rave2006,
       author = {{Steinmetz}, M. and {Zwitter}, T. and {Siebert}, A. and et al.},
        title = "{The Radial Velocity Experiment (RAVE): First Data Release}",
      journal = {\aj},
         year = 2006,
       volume = {132},
        pages = {1645},
          doi = {10.1086/506564}
}

@ARTICLE{2012RAA....12.1197C,
       author = {{Cui}, X.-Q. and {Zhao}, Y.-H. and {Chu}, Y.-Q. and et al.},
        title = "{The Large Sky Area Multi-Object Fiber Spectroscopic Telescope (LAMOST)}",
      journal = {Research in Astronomy and Astrophysics},
         year = 2012,
       volume = {12},
        pages = {1197},
          doi = {10.1088/1674-4527/12/9/003}
}

@ARTICLE{cooper2023,
       author = {{Cooper}, A.~P. and {Koposov}, S.~E. and {Allende Prieto}, C. and et al.},
        title = "{Overview of the DESI Milky Way Survey}",
      journal = {\apj},
         year = 2023,
       volume = {947},
          eid = {37},
          doi = {10.3847/1538-4357/acb3c0}
}

@ARTICLE{dejong2019,
       author = {{de Jong}, R.~S. and {Agertz}, O. and {Berbel}, A.~A. and et al.},
        title = "{4MOST: Project overview and information for the First Call for Proposals}",
      journal = {The Messenger},
         year = 2019,
       volume = {175},
        pages = {3},
          doi = {10.18727/0722-6691/5117}
}

@ARTICLE{2022ApJS..259...35A,
       author = {{Abdurro'uf} and {Accetta}, Katherine and {Aerts}, Conny and {Silva Aguirre}, V{\'\i}ctor and {Ahumada}, Romina and {Ajgaonkar}, Nikhil and {Filiz Ak}, N. and {Alam}, Shadab and {Allende Prieto}, Carlos and {Almeida}, Andr{\'e}s and {Anders}, Friedrich and {Anderson}, Scott F. and {Andrews}, Brett H. and {Anguiano}, Borja and {Aquino-Ort{\'\i}z}, Erik and {Arag{\'o}n-Salamanca}, Alfonso and {Argudo-Fern{\'a}ndez}, Maria and {Ata}, Metin and {Aubert}, Marie and {Avila-Reese}, Vladimir and {Badenes}, Carles and {Barb{\'a}}, Rodolfo H. and {Barger}, Kat and {Barrera-Ballesteros}, Jorge K. and {Beaton}, Rachael L. and {Beers}, Timothy C. and {Belfiore}, Francesco and {Bender}, Chad F. and {Bernardi}, Mariangela and {Bershady}, Matthew A. and {Beutler}, Florian and {Bidin}, Christian Moni and {Bird}, Jonathan C. and {Bizyaev}, Dmitry and {Blanc}, Guillermo A. and {Blanton}, Michael R. and {Boardman}, Nicholas Fraser and {Bolton}, Adam S. and {Boquien}, M{\'e}d{\'e}ric and {Borissova}, Jura and {Bovy}, Jo and {Brandt}, W.~N. and {Brown}, Jordan and {Brownstein}, Joel R. and {Brusa}, Marcella and {Buchner}, Johannes and {Bundy}, Kevin and {Burchett}, Joseph N. and {Bureau}, Martin and {Burgasser}, Adam and {Cabang}, Tuesday K. and {Campbell}, Stephanie and {Cappellari}, Michele and {Carlberg}, Joleen K. and {Wanderley}, F{\'a}bio Carneiro and {Carrera}, Ricardo and {Cash}, Jennifer and {Chen}, Yan-Ping and {Chen}, Wei-Huai and {Cherinka}, Brian and {Chiappini}, Cristina and {Choi}, Peter Doohyun and {Chojnowski}, S. Drew and {Chung}, Haeun and {Clerc}, Nicolas and {Cohen}, Roger E. and {Comerford}, Julia M. and {Comparat}, Johan and {da Costa}, Luiz and {Covey}, Kevin and {Crane}, Jeffrey D. and {Cruz-Gonzalez}, Irene and {Culhane}, Connor and {Cunha}, Katia and {Dai}, Y. Sophia and {Damke}, Guillermo and {Darling}, Jeremy and {Davidson}, Jr., James W. and {Davies}, Roger and {Dawson}, Kyle and {De Lee}, Nathan and {Diamond-Stanic}, Aleksandar M. and {Cano-D{\'\i}az}, Mariana and {S{\'a}nchez}, Helena Dom{\'\i}nguez and {Donor}, John and {Duckworth}, Chris and {Dwelly}, Tom and {Eisenstein}, Daniel J. and {Elsworth}, Yvonne P. and {Emsellem}, Eric and {Eracleous}, Mike and {Escoffier}, Stephanie and {Fan}, Xiaohui and {Farr}, Emily and {Feng}, Shuai and {Fern{\'a}ndez-Trincado}, Jos{\'e} G. and {Feuillet}, Diane and {Filipp}, Andreas and {Fillingham}, Sean P. and {Frinchaboy}, Peter M. and {Fromenteau}, Sebastien and {Galbany}, Llu{\'\i}s and {Garc{\'\i}a}, Rafael A. and {Garc{\'\i}a-Hern{\'a}ndez}, D.~A. and {Ge}, Junqiang and {Geisler}, Doug and {Gelfand}, Joseph and {G{\'e}ron}, Tobias and {Gibson}, Benjamin J. and {Goddy}, Julian and {Godoy-Rivera}, Diego and {Grabowski}, Kathleen and {Green}, Paul J. and {Greener}, Michael and {Grier}, Catherine J. and {Griffith}, Emily and {Guo}, Hong and {Guy}, Julien and {Hadjara}, Massinissa and {Harding}, Paul and {Hasselquist}, Sten and {Hayes}, Christian R. and {Hearty}, Fred and {Hern{\'a}ndez}, Jes{\'u}s and {Hill}, Lewis and {Hogg}, David W. and {Holtzman}, Jon A. and {Horta}, Danny and {Hsieh}, Bau-Ching and {Hsu}, Chin-Hao and {Hsu}, Yun-Hsin and {Huber}, Daniel and {Huertas-Company}, Marc and {Hutchinson}, Brian and {Hwang}, Ho Seong and {Ibarra-Medel}, H{\'e}ctor J. and {Chitham}, Jacob Ider and {Ilha}, Gabriele S. and {Imig}, Julie and {Jaekle}, Will and {Jayasinghe}, Tharindu and {Ji}, Xihan and {Johnson}, Jennifer A. and {Jones}, Amy and {J{\"o}nsson}, Henrik and {Katkov}, Ivan and {Khalatyan}, Dr., Arman and {Kinemuchi}, Karen and {Kisku}, Shobhit and {Knapen}, Johan H. and {Kneib}, Jean-Paul and {Kollmeier}, Juna A. and {Kong}, Miranda and {Kounkel}, Marina and {Kreckel}, Kathryn and {Krishnarao}, Dhanesh and {Lacerna}, Ivan and {Lane}, Richard R. and {Langgin}, Rachel and {Lavender}, Ramon and {Law}, David R. and {Lazarz}, Daniel and {Leung}, Henry W. and {Leung}, Ho-Hin and {Lewis}, Hannah M. and {Li}, Cheng and {Li}, Ran and {Lian}, Jianhui and {Liang}, Fu-Heng and {Lin}, Lihwai and {Lin}, Yen-Ting and {Lin}, Sicheng and {Lintott}, Chris and {Long}, Dan and {Longa-Pe{\~n}a}, Pen{\'e}lope and {L{\'o}pez-Cob{\'a}}, Carlos and {Lu}, Shengdong and {Lundgren}, Britt F. and {Luo}, Yuanze and {Mackereth}, J. Ted and {de la Macorra}, Axel and {Mahadevan}, Suvrath and {Majewski}, Steven R. and {Manchado}, Arturo and {Mandeville}, Travis and {Maraston}, Claudia and {Margalef-Bentabol}, Berta and {Masseron}, Thomas and {Masters}, Karen L. and {Mathur}, Savita and {McDermid}, Richard M. and {Mckay}, Myles and {Merloni}, Andrea and {Merrifield}, Michael and {Meszaros}, Szabolcs and {Miglio}, Andrea and {Di Mille}, Francesco and {Minniti}, Dante and {Minsley}, Rebecca and {Monachesi}, Antonela},
        title = "{The Seventeenth Data Release of the Sloan Digital Sky Surveys: Complete Release of MaNGA, MaStar, and APOGEE-2 Data}",
      journal = {\apjs},
         year = 2022,
        month = apr,
       volume = {259},
       number = {2},
          eid = {35},
        pages = {35},
          doi = {10.3847/1538-4365/ac4414},
archivePrefix = {arXiv},
       eprint = {2112.02026},
 primaryClass = {astro-ph.GA},
       adsurl = {https://ui.adsabs.harvard.edu/abs/2022ApJS..259...35A}
}

@ARTICLE{marcs,
       author = {{Gustafsson}, B. and {Edvardsson}, B. and {Eriksson}, K. and {J{\o}rgensen}, U.~G. and {Nordlund}, {\r{A}}. and {Plez}, B.},
        title = "{A grid of MARCS model atmospheres for late-type stars. I. Methods and general properties}",
      journal = {\aap},
         year = 2008,
        month = aug,
       volume = {486},
       number = {3},
        pages = {951-970},
          doi = {10.1051/0004-6361:200809724},
archivePrefix = {arXiv},
       eprint = {0805.0554},
 primaryClass = {astro-ph},
       adsurl = {https://ui.adsabs.harvard.edu/abs/2008A&A...486..951G}
}

@ARTICLE{2015arXiv150201852H,
       author = {{He}, Kaiming and {Zhang}, Xiangyu and {Ren}, Shaoqing and {Sun}, Jian},
        title = "{Delving Deep into Rectifiers: Surpassing Human-Level Performance on ImageNet Classification}",
      journal = {arXiv e-prints},
         year = 2015,
        month = feb,
          eid = {arXiv:1502.01852},
        pages = {arXiv:1502.01852},
          doi = {10.48550/arXiv.1502.01852},
archivePrefix = {arXiv},
       eprint = {1502.01852},
 primaryClass = {cs.CV},
       adsurl = {https://ui.adsabs.harvard.edu/abs/2015arXiv150201852H}
}

@ARTICLE{2025ApJ...984...58H,
       author = {{Huson}, Dylan and {Cowan}, Indiana and {Sizemore}, Logan and {Kounkel}, Marina and {Hutchinson}, Brian},
        title = "{Gaia Net: Toward Robust Spectroscopic Parameters of Stars of all Evolutionary Stages}",
      journal = {\apj},
         year = 2025,
        month = may,
       volume = {984},
       number = {1},
          eid = {58},
        pages = {58},
          doi = {10.3847/1538-4357/adc2fa},
archivePrefix = {arXiv},
       eprint = {2503.02958},
 primaryClass = {astro-ph.SR},
       adsurl = {https://ui.adsabs.harvard.edu/abs/2025ApJ...984...58H}
}

@ARTICLE{guy2023,
       author = {{Guy}, J. and {Bailey}, S. and {Kremin}, A. and {Alexander}, D.~M. and {Alfarsy}, R. and {Brooks}, D. and {Buckley-Geer}, E. and {Cereskaite}, R. and {Claybaugh}, T. and {Cole}, S. and {Dey}, A. and {Doel}, P. and {Edelstein}, J. and {Eisenstein}, D.~J. and {Font-Ribera}, A. and {Forero-Romero}, J.~E. and {Ho}, S. and {Honscheid}, K. and {Kirkby}, D. and {Kisner}, T. and {Lambert}, A. and {Landriau}, M. and {Levi}, M.~E. and {Meisner}, A. and {Miquel}, R. and {Moustakas}, J. and {Myers}, A.~D. and {Newman}, J.~A. and {Palanque-Delabrouille}, N. and {Prada}, F. and {Sien Tie}, S. and {Schlafly}, E.~F. and {Schlegel}, D. and {Silber}, J.~H. and {Weaver}, B.~A.},
        title = "{The Spectroscopic Data Processing Pipeline for the Dark Energy Spectroscopic Instrument}",
      journal = {\aj},
         year = 2023,
        month = apr,
       volume = {165},
       number = {4},
          eid = {144},
        pages = {144},
          doi = {10.3847/1538-3881/acb212}
}

@ARTICLE{Cirasuolo2020,
       author = {{Cirasuolo}, M. and et al.},
        title = "{MOONS: The New Multi-Object Spectrograph for the VLT}",
      journal = {The Messenger},
         year = 2020,
       volume = {180},
        pages = {10-17},
       eprint = {2009.00628}
}

@ARTICLE{GaiaCollaboration2016,
       author = {{Gaia Collaboration} and {Prusti}, T. and {de Bruijne}, J.~H.~J. and et al.},
        title = "{The Gaia mission}",
      journal = {Astronomy \& Astrophysics},
         year = 2016,
       volume = {595},
          eid = {A1},
          doi = {10.1051/0004-6361/201629272}
}

@ARTICLE{GaiaCollaboration2021,
       author = {{Gaia Collaboration} and {Brown}, A.~G.~A. and {Vallenari}, A. and et al.},
        title = "{Gaia Early Data Release 3. Summary of the contents and survey properties}",
      journal = {Astronomy \& Astrophysics},
         year = 2021,
       volume = {649},
          eid = {A1},
          doi = {10.1051/0004-6361/202039657}
}

@ARTICLE{2023ApJ...950...86G,
       author = {{Gordon}, Karl D. and {Clayton}, Geoffrey C. and {Decleir}, Marjorie and {Fitzpatrick}, E.~L. and {Massa}, Derck and {Misselt}, Karl A. and {Tollerud}, Erik J.},
        title = "{One Relation for All Wavelengths: The Far-ultraviolet to Mid-infrared Milky Way Spectroscopic R(V)-dependent Dust Extinction Relationship}",
      journal = {\apj},
         year = 2023,
        month = jun,
       volume = {950},
       number = {2},
          eid = {86},
        pages = {86},
          doi = {10.3847/1538-4357/accb59},
archivePrefix = {arXiv},
       eprint = {2304.01991},
 primaryClass = {astro-ph.GA},
       adsurl = {https://ui.adsabs.harvard.edu/abs/2023ApJ...950...86G}
}

@ARTICLE{1998A&A...330.1109A,
       author = {{Alvarez}, R. and {Plez}, B.},
        title = "{Near-infrared narrow-band photometry of M-giant and Mira stars: models meet observations}",
      journal = {\aap},
         year = 1998,
        month = feb,
       volume = {330},
        pages = {1109-1119},
          doi = {10.48550/arXiv.astro-ph/9710157},
archivePrefix = {arXiv},
       eprint = {astro-ph/9710157},
 primaryClass = {astro-ph},
       adsurl = {https://ui.adsabs.harvard.edu/abs/1998A&A...330.1109A}
}

@misc{tensorflow2015-whitepaper,
title={ {TensorFlow}: Large-Scale Machine Learning on Heterogeneous Systems},
url={https://www.tensorflow.org/},
note={Software available from tensorflow.org},
author={
    Mart\'{i}n~Abadi and
    Ashish~Agarwal and
    Paul~Barham and
    Eugene~Brevdo and
    Zhifeng~Chen and
    Craig~Citro and
    Greg~S.~Corrado and
    Andy~Davis and
    Jeffrey~Dean and
    Matthieu~Devin and
    Sanjay~Ghemawat and
    Ian~Goodfellow and
    Andrew~Harp and
    Geoffrey~Irving and
    Michael~Isard and
    Yangqing Jia and
    Rafal~Jozefowicz and
    Lukasz~Kaiser and
    Manjunath~Kudlur and
    Josh~Levenberg and
    Dandelion~Man\'{e} and
    Rajat~Monga and
    Sherry~Moore and
    Derek~Murray and
    Chris~Olah and
    Mike~Schuster and
    Jonathon~Shlens and
    Benoit~Steiner and
    Ilya~Sutskever and
    Kunal~Talwar and
    Paul~Tucker and
    Vincent~Vanhoucke and
    Vijay~Vasudevan and
    Fernanda~Vi\'{e}gas and
    Oriol~Vinyals and
    Pete~Warden and
    Martin~Wattenberg and
    Martin~Wicke and
    Yuan~Yu and
    Xiaoqiang~Zheng},
  year={2015},
}

@ARTICLE{2024MNRAS.530.2688J,
       author = {{Jin}, Shoko and {Trager}, Scott C. and {Dalton}, Gavin B. and {Aguerri}, J. Alfonso L. and {Drew}, J.~E. and {Falc{\'o}n-Barroso}, Jes{\'u}s and {G{\"a}nsicke}, Boris T. and {Hill}, Vanessa and {Iovino}, Angela and {Pieri}, Matthew M. and {Poggianti}, Bianca M. and {Smith}, D.~J.~B. and {Vallenari}, Antonella and {Abrams}, Don Carlos and {Aguado}, David S. and {Antoja}, Teresa and {Arag{\'o}n-Salamanca}, Alfonso and {Ascasibar}, Yago and {Babusiaux}, Carine and {Balcells}, Marc and {Barrena}, R. and {Battaglia}, Giuseppina and {Belokurov}, Vasily and {Bensby}, Thomas and {Bonifacio}, Piercarlo and {Bragaglia}, Angela and {Carrasco}, Esperanza and {Carrera}, Ricardo and {Cornwell}, Daniel J. and {Dom{\'\i}nguez-Palmero}, Lilian and {Duncan}, Kenneth J. and {Famaey}, Benoit and {Fari{\~n}a}, Cecilia and {Gonzalez}, Oscar A. and {Guest}, Steve and {Hatch}, Nina A. and {Hess}, Kelley M. and {Hoskin}, Matthew J. and {Irwin}, Mike and {Knapen}, Johan H. and {Koposov}, Sergey E. and {Kuchner}, Ulrike and {Laigle}, Clotilde and {Lewis}, Jim and {Longhetti}, Marcella and {Lucatello}, Sara and {M{\'e}ndez-Abreu}, Jairo and {Mercurio}, Amata and {Molaeinezhad}, Alireza and {Mongui{\'o}}, Maria and {Morrison}, Sean and {Murphy}, David N.~A. and {Peralta de Arriba}, Luis and {P{\'e}rez}, Isabel and {P{\'e}rez-R{\`a}fols}, Ignasi and {Pic{\'o}}, Sergio and {Raddi}, Roberto and {Romero-G{\'o}mez}, Merc{\`e} and {Royer}, Fr{\'e}d{\'e}ric and {Siebert}, Arnaud and {Seabroke}, George M. and {Som}, Debopam and {Terrett}, David and {Thomas}, Guillaume and {Wesson}, Roger and {Worley}, C. Clare and {Alfaro}, Emilio J. and {Allende Prieto}, Carlos and {Alonso-Santiago}, Javier and {Amos}, Nicholas J. and {Ashley}, Richard P. and {Balaguer-N{\'u}{\~n}ez}, Lola and {Balbinot}, Eduardo and {Bellazzini}, Michele and {Benn}, Chris R. and {Berlanas}, Sara R. and {Bernard}, Edouard J. and {Best}, Philip and {Bettoni}, Daniela and {Bianco}, Andrea and {Bishop}, Georgia and {Blomqvist}, Michael and {Boeche}, Corrado and {Bolzonella}, Micol and {Bonoli}, Silvia and {Bosma}, Albert and {Britavskiy}, Nikolay and {Busarello}, Gianni and {Caffau}, Elisabetta and {Cantat-Gaudin}, Tristan and {Castro-Ginard}, Alfred and {Couto}, Guilherme and {Carbajo-Hijarrubia}, Juan and {Carter}, David and {Casamiquela}, Laia and {Conrado}, Ana M. and {Corcho-Caballero}, Pablo and {Costantin}, Luca and {Deason}, Alis and {de Burgos}, Abel and {De Grandi}, Sabrina and {Di Matteo}, Paola and {Dom{\'\i}nguez-G{\'o}mez}, Jes{\'u}s and {Dorda}, Ricardo and {Drake}, Alyssa and {Dutta}, Rajeshwari and {Erkal}, Denis and {Feltzing}, Sofia and {Ferr{\'e}-Mateu}, Anna and {Feuillet}, Diane and {Figueras}, Francesca and {Fossati}, Matteo and {Franciosini}, Elena and {Frasca}, Antonio and {Fumagalli}, Michele and {Gallazzi}, Anna and {Garc{\'\i}a-Benito}, Rub{\'e}n and {Gentile Fusillo}, Nicola and {Gebran}, Marwan and {Gilbert}, James and {Gledhill}, T.~M. and {Gonz{\'a}lez Delgado}, Rosa M. and {Greimel}, Robert and {Guarcello}, Mario Giuseppe and {Guerra}, Jose and {Gullieuszik}, Marco and {Haines}, Christopher P. and {Hardcastle}, Martin J. and {Harris}, Amy and {Haywood}, Misha and {Helmi}, Amina and {Hernandez}, Nauzet and {Herrero}, Artemio and {Hughes}, Sarah and {Ir{\v{s}}i{\v{c}}}, Vid and {Jablonka}, Pascale and {Jarvis}, Matt J. and {Jordi}, Carme and {Kondapally}, Rohit and {Kordopatis}, Georges and {Krogager}, Jens-Kristian and {La Barbera}, Francesco and {Lam}, Man I. and {Larsen}, S{\o}ren S. and {Lemasle}, Bertrand and {Lewis}, Ian J. and {Lhom{\'e}}, Emilie and {Lind}, Karin and {Lodi}, Marcello and {Longobardi}, Alessia and {Lonoce}, Ilaria and {Magrini}, Laura and {Ma{\'\i}z Apell{\'a}niz}, Jes{\'u}s and {Marchal}, Olivier and {Marco}, Amparo and {Martin}, Nicolas F. and {Matsuno}, Tadafumi and {Maurogordato}, Sophie and {Merluzzi}, Paola and {Miralda-Escud{\'e}}, Jordi and {Molinari}, Emilio and {Monari}, Giacomo and {Morelli}, Lorenzo and {Mottram}, Christopher J. and {Naylor}, Tim and {Negueruela}, Ignacio and {O{\~n}orbe}, Jose and {Pancino}, Elena and {Peirani}, S{\'e}bastien and {Peletier}, Reynier F. and {Pozzetti}, Lucia and {Rainer}, Monica and {Ramos}, Pau and {Read}, Shaun C. and {Rossi}, Elena Maria and {R{\"o}ttgering}, Huub J.~A. and {Rubi{\~n}o-Mart{\'\i}n}, Jose Alberto and {Sabater}, Jose and {San Juan}, Jos{\'e} and {Sanna}, Nicoletta and {Schallig}, Ellen and {Schiavon}, Ricardo P. and {Schultheis}, Mathias and {Serra}, Paolo and {Shimwell}, Timothy W. and {Sim{\'o}n-D{\'\i}az}, Sergio and {Smith}, Russell J. and {Sordo}, Rosanna and {Sorini}, Daniele and {Soubiran}, Caroline and {Starkenburg}, Else and {Steele}, Iain A. and {Stott}, John and {Stuik}, Remko and {Tolstoy}, Eline and {Tortora}, Crescenzo and {Tsantaki}, Maria and {Van der Swaelmen}, Mathieu and {van Weeren}, Reinout J. and {Vergani}, Daniela},
        title = "{The wide-field, multiplexed, spectroscopic facility WEAVE: Survey design, overview, and simulated implementation}",
      journal = {\mnras},
         year = 2024,
        month = may,
       volume = {530},
       number = {3},
        pages = {2688-2730},
          doi = {10.1093/mnras/stad557},
archivePrefix = {arXiv},
       eprint = {2212.03981},
 primaryClass = {astro-ph.IM},
       adsurl = {https://ui.adsabs.harvard.edu/abs/2024MNRAS.530.2688J}
}

@ARTICLE{2022ApJS..259...60R,
       author = {{Rockosi}, Constance M. and {Lee}, Young Sun and {Morrison}, Heather L. and {Yanny}, Brian and {Johnson}, Jennifer A. and {Lucatello}, Sara and {Sobeck}, Jennifer and {Beers}, Timothy C. and {Allende Prieto}, Carlos and {An}, Deokkeun and {Bizyaev}, Dmitry and {Blanton}, Michael R. and {Casagrande}, Luca and {Eisenstein}, Daniel J. and {Gould}, Andrew and {Gunn}, James E. and {Harding}, Paul and {Ivans}, Inese I. and {Jacobson}, H.~R. and {Janesh}, William and {Knapp}, Gillian R. and {Kollmeier}, Juna A. and {L{\'e}pine}, S{\'e}bastien and {L{\'o}pez-Corredoira}, Mart{\'\i}n and {Ma}, Zhibo and {Newberg}, Heidi J. and {Pan}, Kaike and {Prchlik}, Jakub and {Sayers}, Conor and {Schlesinger}, Katharine J. and {Simmerer}, Jennifer and {Weinberg}, David H.},
        title = "{SEGUE-2: Old Milky Way Stars Near and Far}",
      journal = {\apjs},
         year = 2022,
        month = apr,
       volume = {259},
       number = {2},
          eid = {60},
        pages = {60},
          doi = {10.3847/1538-4365/ac5323},
       adsurl = {https://ui.adsabs.harvard.edu/abs/2022ApJS..259...60R}
}

@software{turbospectrum,
       author = {{Plez}, B.},
        title = "{Turbospectrum: Code for spectral synthesis}",
 howpublished = {Astrophysics Source Code Library, record ascl:1205.004},
         year = 2012,
        month = may,
          eid = {ascl:1205.004},
       adsurl = {https://ui.adsabs.harvard.edu/abs/2012ascl.soft05004P}
}

@ARTICLE{melissa2015,
       author = {{Ness}, M. and {Hogg}, David W. and {Rix}, H. -W. and {Ho}, Anna. Y.~Q. and {Zasowski}, G.},
        title = "{The Cannon: A data-driven approach to Stellar Label Determination}",
      journal = {\apj},
         year = 2015,
        month = jul,
       volume = {808},
       number = {1},
          eid = {16},
        pages = {16},
          doi = {10.1088/0004-637X/808/1/16},
archivePrefix = {arXiv},
       eprint = {1501.07604},
 primaryClass = {astro-ph.SR},
       adsurl = {https://ui.adsabs.harvard.edu/abs/2015ApJ...808...16N}
}

@ARTICLE{desi2016,
       author = {{DESI Collaboration} and {Aghamousa}, Amir and {Aguilar}, Jessica and {Ahlen}, Steve and {Alam}, Shadab and {Allen}, Lori E. and {Allende Prieto}, Carlos and {Annis}, James and {Bailey}, Stephen et. al.},
        title = "{The DESI Experiment Part I: Science,Targeting, and Survey Design}",
      journal = {arXiv e-prints},
         year = 2016,
        month = oct,
          eid = {arXiv:1611.00036},
        pages = {arXiv:1611.00036},
          doi = {10.48550/arXiv.1611.00036},
archivePrefix = {arXiv},
       eprint = {1611.00036},
 primaryClass = {astro-ph.IM},
       adsurl = {https://ui.adsabs.harvard.edu/abs/2016arXiv161100036D}
}

@ARTICLE{adam2014,
       author = {{Kingma}, Diederik P. and {Ba}, Jimmy},
        title = "{Adam: A Method for Stochastic Optimization}",
      journal = {arXiv e-prints},
         year = 2014,
        month = dec,
          eid = {arXiv:1412.6980},
        pages = {arXiv:1412.6980},
          doi = {10.48550/arXiv.1412.6980},
archivePrefix = {arXiv},
       eprint = {1412.6980},
 primaryClass = {cs.LG},
       adsurl = {https://ui.adsabs.harvard.edu/abs/2014arXiv1412.6980K}
}

@ARTICLE{Payne,
       author = {{Ting}, Yuan-Sen and {Conroy}, Charlie and {Rix}, Hans-Walter and {Cargile}, Phillip},
        title = "{The Payne: Self-consistent ab initio Fitting of Stellar Spectra}",
      journal = {\apj},
         year = 2019,
        month = jul,
       volume = {879},
       number = {2},
          eid = {69},
        pages = {69},
          doi = {10.3847/1538-4357/ab2331},
archivePrefix = {arXiv},
       eprint = {1804.01530},
 primaryClass = {astro-ph.SR},
       adsurl = {https://ui.adsabs.harvard.edu/abs/2019ApJ...879...69T}
}

@ARTICLE{starnet,
       author = {{Fabbro}, S. and {Venn}, K.~A. and {O'Briain}, T. and {Bialek}, S. and {Kielty}, C.~L. and {Jahandar}, F. and {Monty}, S.},
        title = "{An application of deep learning in the analysis of stellar spectra}",
      journal = {\mnras},
         year = 2018,
        month = apr,
       volume = {475},
       number = {3},
        pages = {2978-2993},
          doi = {10.1093/mnras/stx3298},
archivePrefix = {arXiv},
       eprint = {1709.09182},
 primaryClass = {astro-ph.IM},
       adsurl = {https://ui.adsabs.harvard.edu/abs/2018MNRAS.475.2978F}
}

@ARTICLE{ispec2019,
       author = {{Blanco-Cuaresma}, Sergi},
        title = "{Modern stellar spectroscopy caveats}",
      journal = {\mnras},
         year = 2019,
        month = jun,
       volume = {486},
       number = {2},
        pages = {2075-2101},
          doi = {10.1093/mnras/stz549},
archivePrefix = {arXiv},
       eprint = {1902.09558},
 primaryClass = {astro-ph.SR},
       adsurl = {https://ui.adsabs.harvard.edu/abs/2019MNRAS.486.2075B}
}

@ARTICLE{koposov2025,
       author = {{Koposov}, Sergey E. and {Allende Prieto}, Carlos and {Cooper}, Andrew P. et al.},
        title = "{DESI Data Release 1: Stellar Catalogue}",
      journal = {The Open Journal of Astrophysics},
         year = 2026,
        month = jan,
       volume = {9},
          doi = {10.33232/001c.155260},
archivePrefix = {arXiv},
       eprint = {2505.14787},
 primaryClass = {astro-ph.SR},
       adsurl = {https://ui.adsabs.harvard.edu/abs/2025arXiv250514787K}
}

@ARTICLE{zhang2024,
       author = {{Zhang}, Meng and {Xiang}, Maosheng and {Ting}, Yuan-Sen and {Wang}, Jiahui and {Li}, Haining and {Zou}, Hu and {Nie}, Jundan and {Mou}, Lanya and {Wu}, Tianmin and {Wu}, Yaqian and {Liu}, Jifeng},
        title = "{Determining Stellar Elemental Abundances from DESI Spectra with the Data-driven Payne}",
      journal = {\apjs},
         year = 2024,
        month = aug,
       volume = {273},
       number = {2},
          eid = {19},
        pages = {19},
          doi = {10.3847/1538-4365/ad51dd},
archivePrefix = {arXiv},
       eprint = {2402.06242},
 primaryClass = {astro-ph.SR},
       adsurl = {https://ui.adsabs.harvard.edu/abs/2024ApJS..273...19Z}
}

@ARTICLE{vernekar2025,
       author = {{Vernekar}, Nagaraj and {Spina}, Lorenzo and {Lucatello}, Sara and {Arcidiacono}, Carmelo and {Cortese}, Luca and {Simioni}, Matteo and {Balestra}, Andrea},
        title = "{LRPayne: Stellar parameters and abundances from low-resolution spectra}",
      journal = {Astronomy \& Astrophysics, submitted},
         year = 2025,
         note = {manuscript aa56502-25}
}
\bibliographystyle{aasjournal}


\end{document}